\documentclass[a4paper,11pt]{article}
\usepackage{jheppub}
\usepackage{lineno}

\usepackage{tikz}
\usepackage[compat=1.1.0]{tikz-feynman}
\usepackage{bm}
\usepackage{wasysym}

\newcommand{\hats}{\hat{\text{s}}_{W}}
\newcommand{\hatc}{\hat{\text{c}}_{W}}
\newcommand{\sh}{\text{sh} \xi}
\newcommand{\ch}{\text{ch} \xi}

\newcommand{\cbeta}{\text{c}_{\beta}}
\newcommand{\ceta}{\text{c}_{\eta}}
\newcommand{\sbeta}{\text{s}_{\beta}}

\newcommand{\tbeta}{\text{t}_{\beta}}
\newcommand{\teta}{\text{t}_{\eta}}

\usepackage[commentmarkup=uwave]{changes}

\title{\boldmath Electroweak Precision Constraints on Dark Photon Models with Generalized Mixing}

\author[a,b]{Enrico Bertuzzo,}
\author[c]{Csaba Csaki}
\author[c]{and Fernanda Huller}
\affiliation[a]{Dipartimento di Scienze Fisiche, Informatiche e Matematiche, Università degli Studi di Modena e Reggio Emilia, via G. Campi 213/A, 41125 Modena, Italy}
\affiliation[b]{INFN sezione di Bologna, via Irnerio 46, 40126 Bologna, Italy}
\affiliation[c]{Laboratory for Elementary Particle Physics, Cornell University, Ithaca, NY 14853, USA}

\emailAdd{enrico.bertuzzo@unimore.it, csaki@cornell.edu, fh264@cornell.edu}

\abstract{We present a global fit to electroweak precision observables (EWPOs) in dark photon (DP) models containing both kinetic and mass mixing between the DP and the neutral gauge bosons of the Standard Model (SM). Such more general mixing can be the result of an extended scalar sector, which we specify in this paper. We calculate the tree-level contributions to EWPOs due to the mixing with the DP, as well as the leading loop corrections to the oblique parameters due to the extended Higgs sector. In the scalar sector, we find that ample regions of parameter space are still unconstrained by data. In the gauge sector, the excluded region depends strongly on the vacuum expectation values of the scalar fields: for moderate ratios, DP masses in the $(40\,\text{GeV}, 1\,\text{TeV})$ range are excluded; for larger ratios, the limits become indistinguishable from those for standard DPs.}

\begin{document}
\maketitle
\flushbottom

\section{Introduction} \label{sec:intro}

$U(1)$ extensions of the standard model (SM) are ubiquitous in various approaches to beyond the standard model (BSM) physics. Many top-down models contain $Z'$ particles that are expected to be heavy (TeV scale or higher) and mixed with the ordinary $Z$ boson via mass mixing \cite{langacker, carena}. Another set of models that have been widely explored over the past decade are the so called dark photon (DP) models \cite{dark_photons}, motivated to provide novel candidate dark matter (DM) models \cite{fayet}. DPs can be much lighter than the traditional $Z'$ particles from top-down models, as long as their couplings are small \cite{burgess, pospelov, fayet}. In the most commonly considered DP scenarios, the DP mixes with the ordinary gauge bosons via the kinetic mixing term \cite{dark_photons, pospelov, burgess, curtin, Bento:2023flt} 
\begin{equation} \label{eq:kineticmixing}
    \mathcal{L}_{\text{Kinetic}} \supset - \frac{\chi}{2} X_{\mu \nu} B^{\mu \nu},
\end{equation}
where $X_{\mu \nu}$ and $B_{\mu \nu}$ are the DP and hypercharge field strengths, and $\chi$ is the kinetic mixing parameter \cite{holdom}, which is related to the DP-ordinary photon mixing term by $\chi = \epsilon/\cos{\theta_{W}}$\,\cite{dark_photons}. 

The aim of this paper is to explore the simplest scenario where the DP has both kinetic and mass mixings with the neutral SM gauge bosons. In particular, our goal is to establish the bounds on such models from electroweak precision observables (EWPOs) as well as bounds from Higgs physics, as it was also suggested in \cite{Bento:2023flt}. We want to understand whether the addition of the mass mixing changes drastically the experimental limits on the model. The spirit of the analysis is similar to the one presented in\,\cite{Barducci:2021egn,Foguel:2022unm}, in which it was shown that small deformations to the DP model (via the addition of higher dimensional operators of the dark Higgs responsible for the DP mass generation) lead to major changes in the experimental limits.

In order to find a model with both kinetic and mass mixings, the scalar sector of the theory has to be extended. DP models always need to include an additional complex scalar $S$ to break the $U(1)_X$ gauge symmetry and give a mass to the DP \cite{burgess, curtin}. If we tried to use this minimal setup with a single Higgs doublet $H$ and single scalar $S$ to generate both mass and kinetic mixings, we could proceed along the following routes: (i) allowing the complex scalar to have a non-vanishing hypercharge $Y_S\neq 0$ or (ii) giving a non-vanishing $U(1)_X$ charge to the Higgs doublet $H$. In the first case, the electric charge generator is modified to $Q = Q_{SM} - Y_S (x/x_S)$, with $Q_{SM}$ the usual SM generator and $x$ ($x_S$) the $U(1)_X$ charge of the particle under consideration (singlet). However, this relation implies that, in general, the electric charge will not be quantized. In the second scenario, SM fermion masses can be generated only if the SM fermions have a non-vanishing $U(1)_X$ charge. While in principle these options are  possible, they both lead to somewhat unsatisfactory scenarios. 

A simple way to generate both a mass and a kinetic mixing without giving up on charge quantization or being forced to give $U(1)_X$ charges to the SM fermions is to further extend the scalar sector, by adding a second Higgs doublet to the theory \cite{Davoudiasl:2012ag, EricMaximPaper, Cheng:2024hvq, Sun:2023kfu, Campos:2017dgc}. Hence we will have three scalars: a complex singlet $S$ charged under $U(1)_X$ and two SM doublets $H_{1,2}$, with $H_1$ having the quantum numbers of the SM Higgs but uncharged under $U(1)_X$ and $H_2$ being charged under the new Abelian group. It is the vacuum expectation value (vev) of this second doublet that will generate the DP-$Z$ mixing we are interested in, while the vev of $H_1$ will give masses to the SM fermions just as in the SM. Although this model proposes a scalar sector similar to the one in singlet extensions of the minimal supersymmetric standard model (MSSM) and $E_{6}$ supersymmetric Grand Unified Theories (SUSY GUTs) \cite{langacker, Langacker:1998tc, Erler:2000wu, King:2005jy}, we expect it to be phenomenologically different since SM fermions remain uncharged under the extra $U(1)$. Therefore, such simple model will be the focus of our paper.

The paper is organized as follows. In Sec.\,\ref{sec:model} we present the basic structure of our proposed model for DP with kinetic and mass mixing, which is a simple variation of the next-to-minimal two Higgs doublet model (N2HDM) \cite{n2hdm}. In Sec.\,\ref{sec:EWPOs} we show how the expressions for the input parameters of the SM ($\alpha$, $G_F$ and $M_Z$) are modified due to the new interactions with the DP, and provide an expression of the (modified) $Z$ and $Z'$ interactions in terms of the input parameters. Using these results we present expressions for the tree-level corrections to the EWPOs due to the $Z'$. Results from our global fit to the EWPOs can be found in Sec.\,\ref{sec:constraints}. Here we first restrict our parameter space to regions where the contribution of the extended Higgs sector to the $S$, $T$, $U$ oblique parameters is negligible, and find the electroweak precision constraints for this restricted region. We show that the experimental bounds depend crucially on the ratios of vacuum expectation values $\langle H_1 \rangle/\langle H_2\rangle $ and $\langle S \rangle/\langle H_2 \rangle$. More precisely, for moderate values of these ratios ($\lesssim 15-20$) a DP mass in the $(40\,\text{GeV}, 1\,\text{TeV})$ range is excluded. For larger ratios, the contribution to EWPOs coming from the mass mixing is very small and the experimental limits are virtually indistinguishable from those of the usual DP model. Our conclusions are presented in Sec.\,\ref{sec:conclusions}. Further details about the diagonalization of the gauge Lagrangian and the scalar potential can be found in App.\,\ref{app:diagonalization}. Finally, App.\,\ref{app:experimentalValues} contains both SM predictions and experimental values of the observables used in our analysis.

\section{Next-to-Minimal 2HDM with a $U(1)_X$ Symmetry} \label{sec:model} 

As explained in the introduction, we will be considering a model with  $SU(2)_L \times U(1)_Y \times U(1)_X$ gauge group, where $U(1)_X$ will give rise to the DP. Our extended Higgs sector will contain three scalars, with quantum numbers \cite{Davoudiasl:2012ag, EricMaximPaper, Cheng:2024hvq, Sun:2023kfu}
\begin{align}\begin{aligned}\label{eq:quantum_numbers}
H_1 \sim \bm{2}_{1/2, \, 0}, ~~~~~ H_2 \sim \bm{2}_{-1/2, \, -q_X}, ~~~~~ S \sim \bm{1}_{0, \, q_X}.
\end{aligned}\end{align}
$H_1$ will play the role of the usual SM Higgs doublet, providing all fermion masses and most of the SM gauge boson masses. $H_2$ will be responsible for the mass mixing between the $Z$ and $Z'$, while $S$ is the standard scalar needed to generate the $Z'$ mass when a St\"uckelberg term is not simply postulated.  

With the particle content just described, we will consider the Lagrangian
\begin{equation} \label{eq:kineticL}
        \mathcal{L} = \mathcal{L}_{\text{SM}} + \vert D_{\mu} H_{2} \vert^{2} + \vert D_{\mu} S\vert^{2} - \frac{1}{4} \hat{X}_{\mu \nu} \hat{X}^{\mu \nu} - \frac{\chi}{2} \hat{X}_{\mu \nu} \hat{B}^{\mu \nu} - V(H_1, H_2, S),
\end{equation}
where we use a hat to denote gauge fields before kinetic and mass diagonalization. 
The scalar potential is given by \cite{Cheng:2024hvq, Campos:2017dgc}
\begin{equation} \label{eq:potential}
    \begin{aligned}
        V(H_{1}, H_{2}, S) = &- \mu_{1} \vert H_{1}\vert^2 + \lambda_{11} \vert H_{1}\vert^4  - \mu_{2} \vert H_{2}\vert^2 + \lambda_{22} \vert H_{2}\vert^4 - \mu_{S} \vert S\vert^2 + \lambda_{SS} \vert S\vert^4 \\
        & + \lambda_{12} \vert H_{1}\vert^2 \vert H_{2}\vert^2 + \lambda_{21} (H_{1}^{\dagger} H_{2})(H_{2}^{\dagger} H_{1}) + \lambda_{1 S} \vert H_{1}\vert^2 \vert S\vert^2 \\
        & + \lambda_{2 S} \vert H_{2}\vert^2 \vert S\vert^2 + (i \kappa H_{1}^{T} \sigma^{2} H_{2} S + \text{h.c}).
    \end{aligned}
\end{equation}
We assume that the potential parameters are such that all scalar fields acquire a vev,
\begin{equation}
    \langle H_{1,2} \rangle = \frac{v_{1,2}}{\sqrt{2}}, ~~~~ \langle S \rangle = \frac{w}{\sqrt{2}} .
\end{equation}
In Eq.\,\eqref{eq:quantum_numbers} we have chosen opposite $U(1)_X$ charges for $H_2$ and $S$ in order to allow for the cubic term proportional to $\kappa$ in the potential. In its absence, the scalar potential would be invariant under a global $U(1)$ under which all scalar fields transform with the same phase, spontaneously broken by the scalar vevs. A Nambu-Goldstone boson (NGB) would thus appear in the spectrum. An explicit $\kappa \neq 0$ avoids this massless state in the theory.

The presence of such cubic term in Eq.\,(\ref{eq:potential}) can have other important phenomenological consequences. Generally, contributions from an enlarged scalar sector to EWPOs are negligible as long as: (i) the lightest CP-even boson is almost indistinguishable from the SM Higgs; and (ii) all other eigenstates become considerably heavier and degenerate in mass \cite{huntersguide}. However, since $\kappa$ in fact contributes to the mass of all six physical scalar states in our model, it might lead to significant corrections to the $S$, $T$, $U$ oblique parameters.

For future reference, we introduce two combinations of vevs that will be useful in what follows \cite{Campos:2017dgc}:
\begin{equation}\label{eq:def_tangent}
    \tbeta \equiv \tan{\beta} = \frac{v_{1}}{v_{2}}, ~~~~ \teta \equiv \tan{\eta} = \frac{w}{v_{2}}.
\end{equation}

We now turn to the gauge sector. Details on the diagonalization are presented in App.\,\ref{app:gaugeSector}. It is useful to write the physical states (unhatted fields) in terms of the unmixed field $\hat{Z} = \hatc \hat{W^{3}} - \hats \hat{B}$ and $\hat{A} = \hats \hat{W^{3}} + \hatc \hat{B}$, with $\hatc$ and $\hats$ the cosine and sine of the tree-level weak mixing angle (defined in the usual manner). We obtain
\begin{equation} \label{eq:GaugeStates}
    \begin{aligned}
        A &= \hat{A} + \frac{\hatc \, \sh}{\ch} \hat{X}, \\
        Z &= \cos{\theta}\, \hat{Z} + \frac{1}{\ch} \left( \sin{\theta} - \hats \, \sh \, \cos{\theta} \right) \hat{X}, \\
        Z' &= - \sin{\theta}\, \hat{Z} + \frac{1}{\ch} \left( \cos{\theta} + \hats \, \sh \, \sin{\theta}  \right) \hat{X},
    \end{aligned}
\end{equation}
where, following\,\cite{burgess}, we parameterize the kinetic mixing by a hyperbolic sine and cosine as
\begin{equation}
    \sh \equiv  \frac{\chi}{\sqrt{1 - \chi^{2}}}, \quad \text{and} \quad \ch \equiv \frac{1}{\sqrt{1 - \chi^{2}}},
\end{equation}
respectively. The mixing angle $\theta$ between the $Z$ and $Z'$ bosons is
\begin{equation} \label{eq:GaugeMixingAngle}
    \tan{2 \theta} = \frac{2 \, \hats \, \sh - 2 \, \ch \, \cbeta \, \ceta \frac{\hat{M}_{X}}{\hat{M}_{Z}}}{1 - \left(\hats \, \sh\right)^{2} - \left(\ch \frac{\hat{M}_{X}}{\hat{M}_{Z}}\right)^{2} + 2 \, \hats \, \sh \, \ch \, \cbeta \, \ceta \frac{\hat{M}_{X}}{\hat{M}_{Z}}},
\end{equation}
with $c_{\beta, \eta}$ defined as the cosine of the angles corresponding to Eq.\,\eqref{eq:def_tangent} and 
\begin{equation} \label{SMzMass}
    \hat{M}_{Z} \equiv \frac{1}{2} \sqrt{g^{2} + g'^{2}}\,v, ~~~~   \hat{M}_{X} \equiv g_{X} q_{X} \sqrt{w^{2} + v_{2}^{2}} =  \frac{2\,g_X\,q_X}{\sqrt{g^2+g'^2}} \sqrt{\frac{1+\teta^2}{1+\tbeta^2}} \,\hat{M}_Z.
\end{equation}
As usual, we have defined $v^2 \equiv v_1^2 + v_2^2$. Note that the combination $c_\eta\, \hat{M}_X$ is independent of $\teta$. The physical masses of the $Z$ and $Z'$ bosons are easily computed diagonalizing the Lagrangian as shown in App.\,\ref{app:gaugeSector}. In the small mixing angle limit they read
\begin{equation} \label{eq:eigenmasses}
    \begin{aligned}
        M_{Z}^{2} &\approx \hat{M}_{Z}^{2} \left[ 1 + \frac{\hat{M}_{Z}^{2}}{\hat{M}_{Z}^{2} - \hat{M}_{X}^{2}} \left( \hats \, \sh - \frac{1}{\tbeta \, \teta} \frac{\hat{M}_{X}}{\hat{M}_{Z}} \right)^{2} \right], \\
        M_{Z'}^{2} &\approx \hat{M}_{X}^{2} \left[ 1 + \sh^{2} - \frac{1}{(\tbeta \, \teta)^{2}} - \frac{\hat{M}_{Z}^{2}}{\hat{M}_{Z}^{2} - \hat{M}_{X}^{2}} \left( \hats \, \sh - \frac{1}{\tbeta \, \teta} \frac{\hat{M}_{X}}{\hat{M}_{Z}} \right)^{2} \right].
    \end{aligned}
\end{equation}
It is clear from Eqs.\,\eqref{eq:GaugeStates}-\eqref{eq:eigenmasses} that the $Z$ boson mass and couplings are modified with respect to their SM values. Moreover, the $Z'$ boson interacts with SM fermions via $\theta \neq 0$. All this will have important consequences when analyzing the limits that EWPOs can put on the parameter space. We also stress that, as apparent from Eq.\,\eqref{eq:GaugeMixingAngle}, the mixing angle has the usual kinetic mixing contribution proportional to $\sh$, but there are now additional terms that depend on $\tbeta$ and $\teta$. It is the effect of these terms that we will study in the following. The usual DP model is recovered in the $v_2 \to 0$ limit (equivalently, $t_{\beta, \eta} \to \infty$). We will call ``generalized DP'' the case in which $t_{\beta, \eta}$ are finite, and simply ``DP'' the $t_{\beta, \eta} \to \infty$ limit. 

Finally, we briefly discuss the scalar sector of the model. More details can be found in App.\,\ref{app:scalarSector}. Out of the total 10 scalar degrees of freedom, 4 are eaten by the massive vectors and we are thus left with 6 physical scalar states. We have two charged Higgses $H^\pm$ with mass
\begin{equation} \label{eq:ChargedMass}
    M_{H^{\pm}}^{2} = \frac{v}{2} \left( \lambda_{21} \, v + \sqrt{2} \, \kappa \frac{\teta}{\sbeta} \right),
\end{equation}
one pseudoscalar $A^0$ with mass
\begin{equation} \label{eq:OddMass}
    M_{A^{0}}^{2} = \frac{\kappa \, v}{\sqrt{2}} \left( \frac{\sbeta}{\teta} + \frac{\teta}{\sbeta} \right),
\end{equation}
and, finally, three CP-even scalars that we will denote $h^0$, $H^0$ and $s^0$. Before diagonalization, their mass matrix is
\begin{equation} \label{eq:ScalarMatrices}
        M_{\text{CP-even}}^{2} = v^{2} \, \cbeta \begin{pmatrix}
        2 \lambda_{11} \, \sbeta \, \tbeta + \frac{\kappa \, \teta}{\sqrt{2} \, v \, \tbeta} & \lambda_{12} \, \sbeta -  \frac{\kappa \, \teta}{\sqrt{2} \, v} & \lambda_{1S} \, \sbeta \, \teta - \frac{\kappa}{\sqrt{2} \, v} \\
        \lambda_{12} \, \sbeta - \frac{\kappa \, \teta}{\sqrt{2} \, v} & 2 \lambda_{22} \, \cbeta + \frac{\kappa \, \tbeta \, \teta}{\sqrt{2} \, v} & \lambda_{2S} \, \cbeta \, \teta - \frac{\kappa \, \tbeta}{\sqrt{2} \, v} \\
        \lambda_{1S} \, \sbeta \, \teta - \frac{\kappa}{\sqrt{2} \, v} & \lambda_{2S} \, \cbeta \, \teta - \frac{\kappa \, \tbeta}{\sqrt{2} \, v} & 2 \lambda_{SS} \, \cbeta \, \teta^{2} + \frac{\kappa \, \tbeta}{\sqrt{2} \, v \, \teta}
    \end{pmatrix}.
\end{equation}
A few comments are in order. First of all, we see explicitly from Eq.\,\eqref{eq:OddMass} that, in the $\kappa \to 0$ limit, $A^0$ becomes a massless NGB. This is related to the global $U(1)$ symmetry under which the scalar potential is symmetric in this limit, as discussed above. The parameter $\kappa$ also plays other roles: together with $\tbeta$ and $\teta$, it controls the charged scalar mass, as well as the mixing between the CP-even scalars. This has important consequences for the parameter space that we will consider in the following sections. First of all, as already stressed, we want to study the region $t_{\beta,\eta} \sim 5 - 10$ to avoid to reduce the gauge sector of the model to the usual DP one. On the other hand, we need $h_1$ (the radial excitation of $H_1$) to be SM-like, i.e. a relatively small mixing in Eq.\,\eqref{eq:ScalarMatrices}. This points to small-to-moderate values of $\kappa/v$, that in turn imply that $H^\pm$ and $A^0$ cannot be too heavy. We will explore in Sec.\,\ref{sec:fitScalarSector} the values of the scalar sector parameters that are experimentally allowed. 

\section{$Z'$ Contribution to Electroweak Precision Observables} \label{sec:EWPOs}
In order to compute the limits that EWPOs impose on the parameter space of the generalized DP, we will follow the standard procedure: we first express the Lagrangian parameters in terms of input parameters and then use this information to compute all the observables that can be compared with experimental data.
\subsection{Shifting to Physical Parameters} \label{sec:EWparameters}

From Eq.\,(\ref{eq:GaugeStates}) we see that the interactions between the $Z$ boson and SM fermions are modified as \cite{burgess, burgess93}
\begin{equation} \label{eq:ZbosonL}
    \mathcal{L}_{Z} = \frac{\sqrt{4 \pi \hat{\alpha}_{e}}}{\hats \hatc} Z_{\mu} \left[ \bar{\psi}_{L} \gamma^{\mu} \left( g_{\psi L}^{SM} + \delta g_{\psi L} \right) \psi_{L} + \bar{\psi}_{R} \gamma^{\mu} \left( g_{\psi 
    R}^{SM} + \delta g_{\psi R} \right) \psi_{R} \right],
\end{equation}
where $\hat{\alpha}_{e} = \hat{e}^{2}/4 \pi$, $g_{\psi L,R}^{SM}$ are the usual SM couplings and $\delta g_{\psi L,R}$ their deviations due to new physics,
\begin{equation} \label{eq:Zdeviations} 
    \begin{aligned}
        \delta g_{\psi L} &= \left( \cos{\theta} - 1 \right) \left( T_{\psi}^{3} - \hats^{2} Q_{\psi} \right) + \hats \, \sh \, \sin{\theta} \left( T_{\psi}^{3} - Q_{\psi} \right), \\
        \delta g_{\psi R} &= - \left( \cos{\theta} - 1 \right) \hats^{2} Q_{\psi} - \hats \, \sh \, \sin{\theta} \, Q_{\psi}.
    \end{aligned}
\end{equation}
For future reference, we also report the $Z'$ interactions
\begin{equation} \label{eq:ZprimeL}
    \mathcal{L}_{Z'} = \frac{\sqrt{4 \pi \hat{\alpha}_{e}}}{\hats \hatc} Z'_{\mu} \left( \bar{\psi}_{L} \gamma^{\mu} g_{\psi L}^{Z'} \psi_{L} + \bar{\psi}_{R} \gamma^{\mu} g_{\psi R}^{Z'} \psi_{R} \right),
\end{equation}
with couplings given by
\begin{equation} \label{eq:ZprimeCouplings}
    \begin{aligned}
        g_{\psi L}^{Z'} &= - \sin{\theta} \left( T_{\psi}^{3} - \hats^{2} Q_{\psi} \right) + \hats \, \sh \, \cos{\theta} \left( T_{\psi}^{3} - Q_{\psi} \right), \\
        g_{\psi R}^{Z'} &= (\sin{\theta} \, \hats^{2}   - \hats \, \sh \, \cos{\theta}) \, Q_{\psi}.
    \end{aligned}
\end{equation}
We now need to eliminate the ($\hat{e}$, $\hat{M}_{Z}$, $\hat{\theta}_{W}$) parameters that appear in the tree-level Lagrangian in favor of reference measured quantities \cite{Schwartz}. For this purpose, we follow the standard choice and consider the best-measured observables as input: $\alpha_{e}$, $M_{Z}$, and $G_{F}$ \cite{burgess93, Schwartz}. All expressions can be linearized in terms of the extra contributions given that they are expected to be small \cite{burgess93}.
\begin{itemize}
    \item \textbf{Fine-Structure Constant.} The anomalous magnetic moment of the
    electron \cite{measure_alpha} still remains the most reliable way of extracting $\alpha_{e}$ for the purpose of electroweak physics. Since the photon coupling to fermions is not modified by the presence of the DP, the main modification to the $g-2$ of any lepton comes from a triangle one-loop diagram with exchange of a virtual $Z'$ boson, with couplings given by Eq.\,(\ref{eq:ZprimeL}). At leading order in the $Z'$ corrections the loop gives
\begin{equation} \label{eq:finalalpha}
    \begin{split}
    \hat{\alpha}_{e}& \approx \alpha_{e}(0) \\
    &- \frac{ m_{e}^{2} \, \alpha_{e}(0)}{2 \, \hats^{2} \hatc^{2}} \int_{0}^{1} dz \frac{(g_{e V}^{Z'})^{2} z (1-z)^{2} - (g_{e A}^{Z'})^{2} \left[ z (1-z) (z+3) + z (1-z)^{3} \frac{m_{e}^{2}}{M_{Z'}^{2}} \right]}{(1-z)^{2} m_{e}^{2} + z M_{Z'}^{2}},
    \end{split}
\end{equation}
with $g_{e V,A}^{Z'} = g_{e L}^{Z'} \pm g_{e R}^{Z'}$ the vector and axial couplings, respectively. Although it may seem that the DP correction diverges in the $M_{Z'} \rightarrow 0$ limit, in reality this divergence is compensated by the fact that, for light $Z'$, $g_{e A}^{Z'} \propto M_{Z'}$, so that $\hat{\alpha}_{e}$ is always finite. This is consistent with the fact that a massless gauge boson must couple to a conserved current. 

The value of the electromagnetic coupling extracted from the $g-2$ of the electron is taken near the Thomson limit \cite{pdg}. For precision electroweak physics, however, most observables are measured at and beyond the $Z$ resonance, so that it would be convenient to use $\alpha_{e}(M_{Z})^{-1} = 127.951(9)$ \cite{pdg} as input parameter. This can be done by including radiative corrections directly into $\hat{\alpha}_e$ and substituting $\alpha_e(0) \to \alpha_e(M_Z)$ in Eq.\,\eqref{eq:finalalpha}.

\item \textbf{$Z$ boson mass.} As we saw in Eq.\,\eqref{eq:eigenmasses}, in our model the physical $Z$ boson mass, $M_{Z} = 91.1876(21)$ GeV \cite{pdg}, differs from the SM prediction. It is useful to parametrize the difference as \cite{burgess,burgess93}
\begin{equation} \label{eq:correctionzmass}
        z \equiv \frac{(M_{Z}^{2} - \hat{M}_{Z}^{2})}{\hat{M}_{Z}^{2}}.
\end{equation}
This is one of the important quantities to appear in the calculation of observables.

Moreover, in order to completely eliminate $\hat{M}_{Z}$ in favor of the observable $Z$ mass, we must invert the mass eigenvalue expressions of Eq.\,(\ref{eq:eigenmasses}). In the limit of small mixing, it is then straightforward to obtain the parameter $z$ exclusively in terms of the physical masses as
\begin{equation}
    z \approx \frac{M_{Z}^{2}}{M_{Z}^{2} - M_{Z'}^{2}} \left( \hats \, \sh - \frac{1}{\tbeta \, \teta} \frac{M_{Z'}}{M_{Z}} \right)^{2}.
\end{equation}
\item \textbf{Fermi constant.} The most precise experimental determination of the Fermi constant to date has been through the measurement of the muon lifetime: $G_{F} = 1.1663788(6) \times 10^{-5}$ GeV \cite{pdg}. Since our model does not alter the charged gauge sector, the Fermi constant is simply given by the usual formula,
\begin{equation} \label{eq:FermiConstant}
    G_{F} = \frac{\pi \, \alpha_{e}(0)}{\sqrt{2} \, s_{W}^{2} c_{W}^{2} M_{Z}^{2}}.
\end{equation}
Plugging Eqs.\,(\ref{eq:finalalpha}) and (\ref{eq:correctionzmass}) into the expression above and comparing it with the SM tree-level prediction, we can infer how the weak mixing angle is modified in the presence of a DP. In terms of only observable quantities, the result is given by
\begin{equation} \label{eq:weakcosine}
    \begin{gathered}
    \hats^{2} = s_{W}^{2} \left[ 1 + \frac{c_{W}^{2}}{\left( c_{W}^{2} - s_{W}^{2} \right)} \left( - \Delta \alpha_{e} + z \right) \right], \\ \hatc^{2} = c_{W}^{2} \left[ 1 - \frac{s_{W}^{2}}{\left( c_{W}^{2} - s_{W}^{2} \right)} \left( - \Delta \alpha_{e} + z \right) \right],
    \end{gathered}
\end{equation}
where we have parameterized the correction to the fine-structure constant as $\alpha_{e} (0) = \hat{\alpha}_{e} (1 + \Delta \alpha_{e})$. Once more, when considering precision tests it is convenient to connect Eq.\,(\ref{eq:FermiConstant}) to $\alpha_{e}(M_{Z})$ by the running of the electromagnetic coupling \cite{Schwartz}. The same can be done for the weak mixing angle, so that we will use $s_W^2 = s_{W}^{2}(M_{Z}) \approx 0.23$ \cite{pdg} including radiative corrections in $\hat{s}_W$ and $\hat{c}_W$. 
\end{itemize}

We can finally re-express the interactions of Eqs.\,(\ref{eq:ZbosonL}) and (\ref{eq:ZprimeL}) in terms of the standard electroweak parameters. For the $Z$ boson couplings we write
\begin{equation} \label{eq:Zcouplings}
    \mathcal{L}_{Z} = \frac{\sqrt{4 \pi \alpha_{e}(M_{Z})}}{s_{W} c_{W}} Z_{\mu} \left[ \bar{\psi}_{L} \gamma^{\mu} \left( g_{\psi L}^{SM} + \Delta g_{\psi L} \right) \psi_{L} + \bar{\psi}_{R} \gamma^{\mu} \left( g_{\psi R}^{SM} + \Delta g_{\psi R} \right) \psi_{R} \right].
\end{equation}
The shifts to the SM couplings are now given by 
\begin{equation}
    \Delta g_{\psi L,R} = \delta g_{\psi L,R} - z \left[ \frac{1}{2} g_{\psi L, R}^{SM} + \frac{s_{W}^{2} c_{W}^{2}}{\left( c_{W}^{2} - s_{W}^{2} \right)} Q_{\psi} \right],
\end{equation}
with the SM couplings themselves given in terms of unhatted quantities. In addition, the Lagrangian that describes the $Z'$ boson interactions now becomes  
\begin{equation} 
    \mathcal{L}_{Z'} = \frac{\sqrt{4 \pi \alpha_{e}(M_{Z})}}{s_{W} c_{W}} Z'_{\mu} \left( \bar{\psi}_{L} \gamma^{\mu} g_{\psi L}^{Z'} \psi_{L} + \bar{\psi}_{R} \gamma^{\mu} g_{\psi R}^{Z'} \psi_{R} \right).
\end{equation}
Since they are already small, the direct couplings between the DP and the SM fermions remain the same as in Eq.\,(\ref{eq:ZprimeCouplings}).

\subsection{Calculation of Observables} \label{sec:observables}
According to the usual global analysis performed to derive constraints on distinct DP models \cite{EricMaximPaper, Cheng:2024hvq, Sun:2023kfu, langacker, burgess, curtin, Barbieri:2004qk, Cacciapaglia:2006pk}, we consider the following observables: (i) EWPOs measured at the $Z$-peak by LEP1 and SLC \cite{lep1}, (ii) differential cross sections measured by LEP2 at center of mass energies above $M_{Z}$ \cite{lep2}, (iii) the $W$ mass determined by the ATLAS collaboration \cite{atlas}, (iv) low-energy neutrino-electron scattering measured by the CHARM-II collaboration \cite{charm2}, and (v) electron and muon $g-2$ \cite{pospelov, burgess}. Experimental values and SM predictions for the variables considered can be found in App.\,\ref{app:experimentalValues}.

Since we anticipate new physics effects to be small, in all the expressions derived below they can be considered only at leading order. Moreover, for observables measured at the $Z$ peak, we expect the DP contribution and the $Z - Z'$ interference to give just subdominant contributions -- unless $M_{Z'} \simeq M_Z$, which is a case that we already know must be excluded. 

We can further show that at both $Z$ resonance and beyond the shift to the $Z$ boson mass dominates over the correction to the fine-structure constant, meaning that $\Delta \alpha_{e}$ is only relevant for low-energy processes. As a consequence, we neglect the contribution from $\Delta \alpha_{e}$ to Eq.\,(\ref{eq:weakcosine}) when deriving high-energy constraints, and directly use the accuracy of $g-2$ measurements to impose bounds on our model. 

\begin{itemize}
\item \textbf{High-Energy Observables.}
We start with a set of observables measured at the $Z$ peak by LEP1 and SLC, which can be related to the $Z$ boson partial widths into fermion-antifermion pairs $\psi\bar{\psi}$. Using Eq.\,\eqref{eq:Zcouplings}, this width can be written as
\begin{equation} \label{eq:decaywidth}
    \begin{aligned}
        \Gamma_{\psi} &= \Gamma_{\psi}^{SM} \left[ 1 + \frac{2\,g_{\psi L}^{SM}}{(g_{\psi L}^{SM})^{2} + (g_{\psi R}^{SM})^{2}} \, \Delta g_{\psi L} + \frac{2\,g_{\psi R}^{SM}}{(g_{\psi L}^{SM})^{2} + (g_{\psi R}^{SM})^{2}} \, \Delta g_{\psi R} \right] \\
        &\equiv \Gamma_{\psi}^{SM} (1 + \delta \Gamma_{\psi}),
    \end{aligned}
\end{equation}
where the correction $\delta \Gamma_{\psi}$ is defined to be dimensionless. The observables that we will consider are (i) the total $Z$ decay width $\Gamma_Z \equiv \sum_\psi \Gamma_\psi$, (ii) the total annihilation cross section into hadrons $\sigma_\text{had} \equiv 12\pi \Gamma_e \,\Gamma_\text{had}/(M_Z^2 \Gamma_Z^2)$ (with $\Gamma_\text{had}$ the total decay width into quarks), (iii) the ratio of partial widths into leptons $R_l \equiv \Gamma_\text{had}/\Gamma_l$ and into quarks $R_q \equiv \Gamma_q/\Gamma_\text{had}$\,\cite{feldman}. In terms of the various $\delta\Gamma_\psi$, they read
\begin{align}
       \Gamma_{Z} &= \Gamma_{Z}^{SM} + 3 \left( \frac{2}{3} \Gamma_{u}^{SM} \delta \Gamma_{u} + \Gamma_{d}^{SM} \delta \Gamma_{d} + \Gamma_{e}^{SM} \delta \Gamma_{e} + \Gamma_{\nu}^{SM} \delta \Gamma_{\nu} \right), \nonumber \\
        \begin{split} \sigma_{\text{had}} &= \sigma_{\text{had}}^{SM} \left[ 1 + \left( \frac{1}{\Gamma_{\text{had}}^{SM}} - \frac{2}{\Gamma_{Z}^{SM}} \right) \right. \left( 2 \Gamma_{u}^{SM} \delta \Gamma_{u} + 3 \Gamma_{d}^{SM} \delta \Gamma_{d} \right) \\ 
        &\phantom{ = {}} + \left. \left( 1 - 6 \frac{\Gamma_{e}^{SM}}{\Gamma_{Z}^{SM}} \right) \delta \Gamma_{e} - 6 \frac{\Gamma_{\nu}^{SM}}{\Gamma_{Z}^{SM}} \delta \Gamma_{\nu} \right], \end{split} \\
        R_{l} &= R_{l}^{SM} \left[ 1 + \frac{2 \Gamma_{u}^{SM} \delta \Gamma_{u} + 3 \Gamma_{d}^{SM} \delta \Gamma_{d}}{\Gamma_{\text{had}}^{SM}}  - \delta \Gamma_{l} \right], \nonumber \\
        R_{q} &= R_{q}^{SM} \left[ 1 - \frac{2 \Gamma_{u}^{SM} \delta \Gamma_{u} + 3 \Gamma_{d}^{SM} \delta \Gamma_{d}}{\Gamma_{\text{had}}^{SM}}  + \delta \Gamma_{q} \right]. \nonumber
\end{align} 
Important constraints also follow from asymmetry parameters at the $Z$ pole. We will focus on four such observables: the electron polarization (or left-right) asymmetry $\mathcal{A}^{e}_{LR} \equiv \mathcal{A}_{e}$, the forward-backward asymmetry for quarks and leptons $\mathcal{A}_{FB}^{q,l} \equiv 3 \mathcal{A}_e \mathcal{A}_{q,l}/4$, the forward-backward polarization asymmetry for $\tau$ leptons in the final state $\mathcal{A}_{e} (P_{\tau}) \equiv \mathcal{A}_{e}$, and the average final-state $\tau$ polarization $\mathcal{A}_{\tau} (P_{\tau}) \equiv \mathcal{A}_{e}$\footnote{Given that our model respects lepton universality, the expressions for $\mathcal{A}_{e, \tau} (P_{\tau})$ are the same as the one for $A_{LR}^{e}$. However, since they were measured independently at LEP, we include all three observables in our analysis.}. These observables are computed in terms of 
\begin{equation}
    \mathcal{A}_{\psi} = \frac{g_{\psi L}^{2} - g_{\psi R}^{2} }{ g_{\psi L}^{2} + g_{\psi R}^{2} } = \mathcal{A}_{\psi}^{SM} + \delta \mathcal{A}_{\psi},
\end{equation}
where, using Eq.\,\eqref{eq:Zcouplings} once more, the NP contribution can be written as
\begin{equation} \label{eq:asymmetry}
        \delta \mathcal{A}_{\psi} =  \frac{4\,g_{\psi L}^{SM} \, g_{\psi R}^{SM}}{\left((g_{\psi L}^{SM})^{2} + (g_{\psi R}^{SM})^{2}\right)^{2}} \left(g_{\psi R}^{SM} \Delta g_{\psi L} - g_{\psi L}^{SM} \Delta g_{\psi R}\right) .
\end{equation}
Linearizing in the $\delta\mathcal{A}_\psi$'s, we find
\begin{equation}
    \begin{aligned}
    \mathcal{A}_{LR}^{e} &= \mathcal{A}_{e, \tau} (P_{\tau}) = \mathcal{A}_{e}^{SM} + \delta \mathcal{A}_{e}, \\
    \mathcal{A}_{FB}^{l} &= \frac{3}{4} \left( \mathcal{A}_{e}^{SM} \mathcal{A}_{l}^{SM} + 2 \mathcal{A}_{e}^{SM} \delta \mathcal{A}_{l} \right), \\
    \mathcal{A}_{FB}^{q} &= \frac{3}{4} \left( 1 - k_{A} \frac{\alpha_{s}}{\pi} \right) \left[ \mathcal{A}_{e}^{SM} \mathcal{A}_{q}^{SM} + \left( \mathcal{A}_{e}^{SM} \delta \mathcal{A}_{q} + \mathcal{A}_{q}^{SM} \delta \mathcal{A}_{e} \right) \right],
    \end{aligned}
\end{equation}
where the factor $1 - k_{A} \frac{\alpha_{s}}{\pi}\simeq 0.93$ comes from QCD radiative corrections\,\cite{burgess93}.

We also consider observables measured above the $Z$ pole by LEP2, with energies from $130$ GeV to around $209$ GeV. In particular, we focus on the total production cross section of fermion-antifermion pairs for $\psi = \mu$, $\tau$, and $q$ (hadrons) \cite{pdg}, for which we have 
\begin{equation} \label{eq:lep2cross}
    \begin{split}
        \sigma (e^{+} e^{-} \rightarrow \bar{\psi} \psi) = N_{c} & \frac{\pi \, s \, \alpha_{e}^{2}(M_{Z})}{3} \bigg[ |A_{LL}|^{2} + |A_{LR}|^{2} + |A_{RL}|^{2} + |A_{RR}|^{2} \bigg], \\
    \end{split}
\end{equation}
where $N_c = 1 \, (3)$ for leptons (quarks), $\sqrt{s}$ is the center-of-mass energy, and $|A_{IJ}|^{2}$ ($I,J = L,R$) is the square of the helicity amplitude, defined as
\begin{equation} \label{eq:helicity}
    \begin{split}
        |A_{IJ}|^{2} = (&A_{IJ}^{SM})^{2} \\
        & + \frac{2 A_{IJ}^{SM}}{s_{W}^{2} c_{W}^{2}} \left[ \frac{(s - M_{Z'}^{2}) \, g_{e I}^{Z'} g_{\psi J}^{Z'}}{(s - M_{Z'}^{2})^{2} + (\Gamma_{Z'} M_{Z'})^{2}} + \frac{(g_{e I}^{SM} \Delta g_{\psi J} + g_{\psi J}^{SM} \Delta g_{e I})}{s - M_{Z}^{2}} \right],
    \end{split}
\end{equation}
with the SM helicity amplitude given by
\begin{equation}
    A_{IJ}^{SM}(s) = \frac{Q_{e} Q_{\psi}}{s} + \frac{1}{s_{W}^{2} c_{W}^{2}} \frac{ g_{e I}^{SM} \, g_{\psi J}^{SM}}{s - M_{Z}^{2}} .
\end{equation}
These expressions have been derived in the $\sqrt{s} \gg m_{e,\psi}$ limit. Moreover, we have neglected the term resulting from the axial coupling of the $Z'$ to fermions. This term would be proportional to $(m_e\,m_\psi/M_{Z'}^2)g_{e A}^{Z'} g_{\psi A}^{Z'} $ and, apparently, could be enhanced in the $m_{e, \psi}/M_{Z'} \gg 1$ limit. This however does not happen: as we have seen in our discussion of Eq.\,\eqref{eq:finalalpha}, for small $Z'$ masses the axial coupling is proportional to $M_{Z'}$ and this term ends up being negligible. 

The final non-$Z$ pole observable we consider is the $W$ boson mass. Although the charged gauge Lagrangian is not directly modified in the presence of the extra gauge boson, the SM mixing with the $Z'$ modifies the prediction for the $W$ mass. At lowest order in the electroweak theory, $M_{W}$ can be expressed solely as a function of the $Z$ boson mass and the weak mixing angle. Thus, the expression we must use in our analysis is easily determined by plugging Eqs.\,(\ref{eq:correctionzmass}) and (\ref{eq:weakcosine}) into the SM prediction $\hat{M}_{W}^{2} = \hatc^{2} \hat{M}_{Z}^{2}$:
\begin{equation}
    M_{W} = M_{W}^{SM} \left[ 1 - \frac{c_{W}^{2}}{2 (c_{W}^{2} - s_{W}^{2})} z \right],
\end{equation}
where $M_{W}^{SM}$ can be computed at any desired loop order in the SM interactions \cite{burgess93}. 

\item\textbf{Intermediate-Energy Observables.} 
In order to impose constraints on lighter $Z'$ bosons, we extend our analysis by including low-energy scattering of muon-neutrinos with electrons \cite{burgess}. While studying the differential cross section of such process, the CHARM-II collaboration at CERN was able to precisely determine $s_{W}^{2}$ through the measurement of the cross section ratio \cite{charm2}
\begin{equation}
    R = \frac{\sigma (\nu_{\mu} e^{-} \rightarrow \nu_{\mu} e^{-})}{\sigma (\bar{\nu}_{\mu} e^{-} \rightarrow \bar{\nu}_{\mu} e^{-})}.
\end{equation} 
However, since the direct extraction of this quantity in our model is not trivial, we reconstructed the scattering cross section (and, consequently, the ratio $R$) using the effective vector and axial coupling constants measured by the collaboration, as well as the SM predictions given by PDG -- see Table\,\ref{tab:measurementsCHARM} in App.\,\ref{app:experimentalValues}.

Since both $\nu_{\mu} e^{-}$ and $\bar{\nu}_{\mu} e^{-}$ scattering can now be mediated by a DP, their total cross section is also substantially modified. Considering only left-handed neutrinos and remembering that the $Z$ boson mass is much greater than the invariant energy exchanged in the process ($\sqrt{t} \ll M_{Z}$), we can easily determine an expression for the differential cross section which accounts for the exchange of a $Z'$ boson. In the case of neutrinos with a mean energy of $E_{\nu} = 23.8$ GeV in the LAB frame\,\cite{2charm2}, after integration over phase space we obtain
\begin{equation} \label{eq:scattering}
    \begin{split}
        \sigma (\nu_{\mu} e^{-} \rightarrow \nu_{\mu} e^{-})& = \frac{2 G_{F}^{2} m_{e} E_{\nu}}{\pi} \\
        & \times \left\{ \frac{4}{3} (g_{\nu L}^{SM})^{2} \left[ 3 (g_{e L}^{SM})^{2} + (g_{e R}^{SM})^{2} \right] + \Delta\sigma^{Z} + \Delta\sigma^{Z'} \right\},
    \end{split}
\end{equation}
where the corrections $\Delta \sigma^{Z}$ and $\Delta \sigma^{Z'}$ indicate the contributions due to changes in the $Z$ boson couplings and $Z -Z'$ interference, respectively. Their explicit expressions are
\begin{equation} \label{eq:scatteringcorrection}
    \begin{split}
        &\begin{split}
            \Delta & \sigma^{Z} = \frac{8}{3} g_{\nu L}^{SM} \\
            & \times \left\{ \delta g_{\nu L} \left[ 3 (g_{e L}^{SM})^{2} + (g_{e R}^{SM})^{2} \right] + g_{\nu L}^{SM} \left( 3 g_{e L}^{SM} \delta g_{e L} + g_{e R}^{SM} \delta g_{e R} \right) \right\},
        \end{split} \\
        &\begin{split}
            \Delta & \sigma^{Z'} = \frac{M_{Z}^{2}}{m_{e} E_{\nu}} g_{\nu L}^{SM} g_{\nu L}^{Z'} \, g_{e R}^{SM} g_{e R}^{Z'} \\ 
            & \times \left\{ 4 \frac{g_{e L}^{SM} g_{e L}^{Z'}}{g_{e R}^{SM} g_{e R}^{Z'}} \, \log \left( \frac{2 m_{e} E_{\nu}}{M_{Z'}^{2}} + 1 \right) - 2 \left[ \log \left( \frac{M_{Z'}^{4}}{(2 m_{e} E_{\nu} + M_{Z'}^{2})^{2}} \right) + 3 \right] \right. \\
            & \left. - \frac{2 M_{Z'}^{2}}{m_{e} E_{\nu}} \left[ \log \left( \frac{M_{Z'}^{4}}{(2 m_{e} E_{\nu} + M_{Z'}^{2})^{2}} \right) + 1 \right] - \frac{M_{Z'}^{4}}{m_{e}^{2} E_{\nu}^{2}} \log \left( \frac{M_{Z'}^{2}}{2 m_{e} E_{\nu} + M_{Z'}^{2}} \right) \right\}.
        \end{split}
    \end{split}
\end{equation}
We have kept only terms to the lower order in the NP parameters. 
The cross section for the scattering of antineutrinos can be found by substituting $g_{eL,R}^{SM} \leftrightarrow g_{eR,L}^{SM}$ in Eqs.\,(\ref{eq:scattering}) and (\ref{eq:scatteringcorrection}), and using $E_{\bar{\nu}} = 19.3$ GeV \cite{2charm2}. 

\item\textbf{Low-Energy Observables.} 
As discussed in Sec.\,\ref{sec:EWparameters}, measurements from low-energy processes like $(g-2)_{e}$ can be used to impose stringent constraints on extra gauge bosons with sub-GeV masses \cite{burgess, pospelov}. The next most precise way of determining the electron's anomalous magnetic moment comes from atomic physics results, which leads to a bound on the $Z'$ correction of $\delta a_{e} < 1.59 \times 10^{-10}$ \cite{burgess, pospelov}, where we have defined $\delta a_{e} = (g-2)_{e}^{Z'}/2$. The measurement of the muon $g-2$ provides an additional constraint on the parameters of a DP. Following the analysis of \cite{pospelov}, the $Z'$ correction can be further limited to satisfy $\delta a_{\mu} < 7.4 \times 10^{-9}$. As it can be read off from Eq.\,\eqref{eq:finalalpha}, the two observables depend on $m_l/M_{Z'}$ (with $l =e, \mu$), such that we expect the two observables to limit complementary regions of parameter space.
\end{itemize}

\section{Constraints on the Parameter Space} \label{sec:constraints}

As we have seen in Sec.\,\ref{sec:model}, the model we are considering not only modifies the gauge sector, but also introduces new scalar fields. Except for the SM-like Higgs boson $h^{0}$, the additional scalars will have suppressed couplings to fermions (this coupling is proportional to the mixing with the SM-like state, which we know must be small). However, they can have unsuppressed couplings to gauge bosons. This is the case for the states that emerge mainly from the doublet $H_2$, i.e. the pseudoscalar $A^0$, the charged Higgs $H^\pm$, and one of the CP-even scalars, denoted here by $H^{0}$. In the limit of small \textit{scalar} mixing, we can easily verify that couplings involving the singlet-like Higgs $s^{0}$ can be orders of magnitude smaller than all other gauge-scalar couplings. Consider, for example, the Higgs-vector-vector couplings. In order to avoid reducing our gauge sector to the usual DP model, we are interested in the region of moderate $\tbeta, \teta$ for which $H^{0}$ would have somewhat enhanced couplings. Since in models with more complicated Higgs sectors the CP-even bosons satisfy sum rules as \cite{huntersguide}
\begin{equation}
    g_{h^{0} V V}^{2} + g_{H^{0} V V}^{2} + g_{s^{0} V V}^{2} = g_{h^{SM} V V}^{2},
\end{equation}
it follows that $s^{0}$ must have the most suppressed couplings. On the other hand, in the limit of small \textit{gauge} mixing, couplings to the DP will also be highly suppressed, regardless of the scalar state involved.

The states emerging from the extra doublet can, therefore, contribute to the gauge boson vacuum polarizations and affect observables in a way that is not captured by our discussion in Sec.\,\ref{sec:observables}, which focused on the $Z'$ contributions. Before analyzing the limits on the $Z'$ parameter space, we will further determine whether a region in parameter space exists where the contributions of the NP scalars to observables can be neglected. We will contend ourselves with a simplified analysis in which we will consider bounds coming from oblique parameters. Since, as already mentioned, the only states that can significantly contribute are those coming from the additional doublet, we will use the results for 2HDMs described in \cite{rhoParameter, StUparameters}. One crucial point is that large contributions to the oblique parameters appear when the scalar masses are very different.

\subsection{Bounds on the Scalar Sector} \label{sec:fitScalarSector}

\begin{figure}[t]
    \centering
    \includegraphics[width=0.49\textwidth]{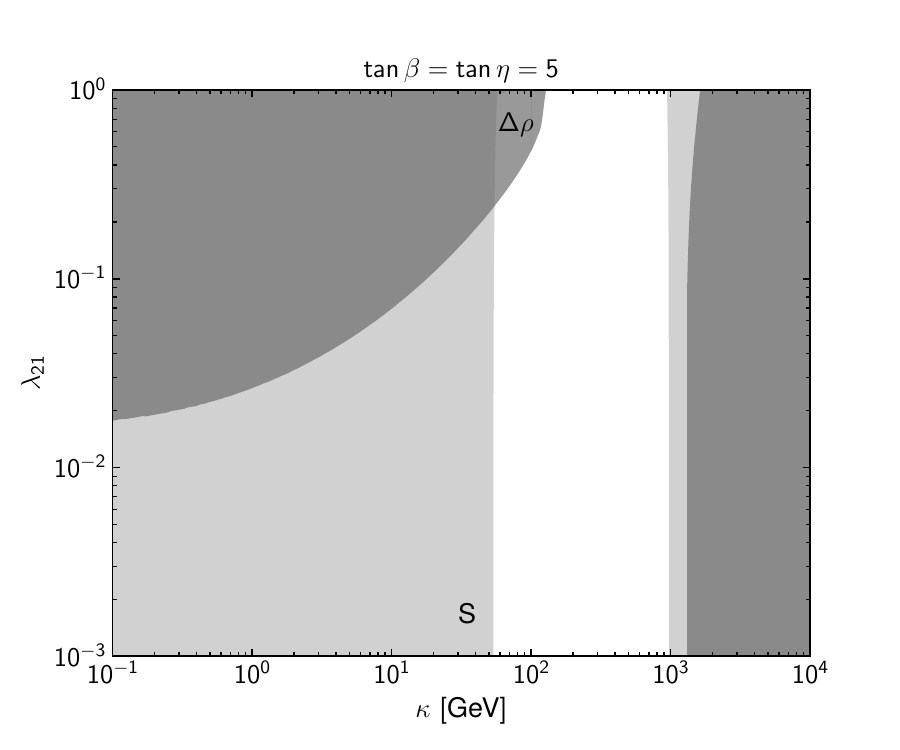}
    \includegraphics[width=0.49\textwidth]{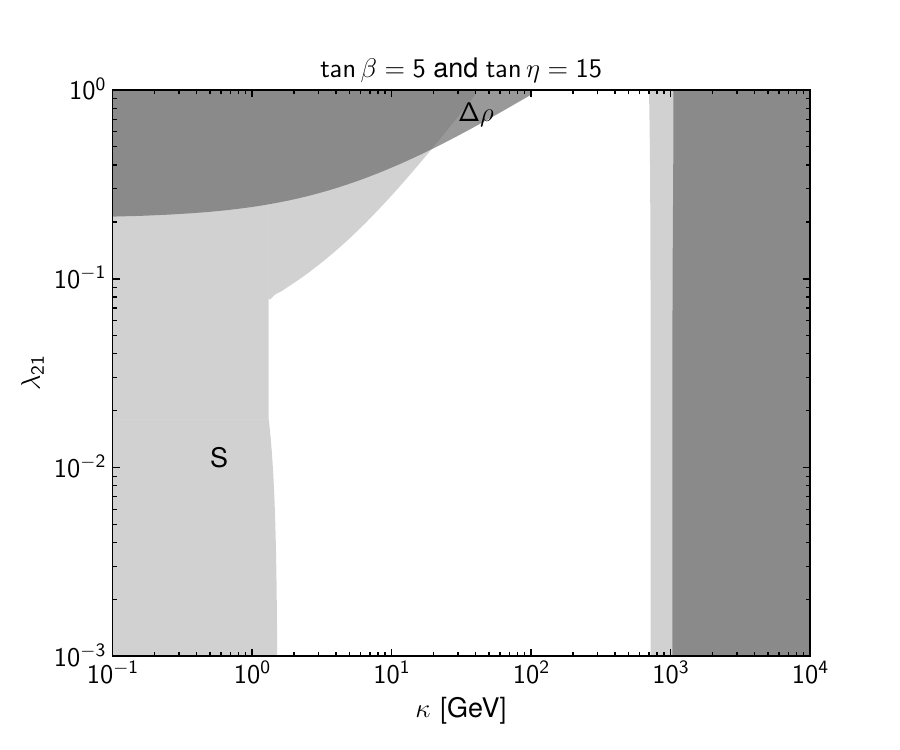}
    \caption{Scalar parameter space excluded by $S$ (light gray region) and $\Delta\rho$ (dark gray region). The constraints have been computed requiring the scalar contributions to $S$ ($\Delta\rho$) to be smaller than $10^{-2}$ ($10^{-4}$). Small values of $\kappa$ are excluded because in this region the scalar spectrum is highly non-degenerate. Large values of $\kappa$ are excluded because the SM-like Higgs mass deviates from the current experimental value.
    }
    \label{fig:scalar_sector_oblique_parameters}
\end{figure}

Electroweak data allows us to determine the oblique parameters \cite{pdg}
\begin{equation} \label{eq:experimentalRHOandS}
        S = -0.5 \pm 0.07, ~~~~~~~~ \Delta \rho = \alpha_{e} T = 1.0000 \pm 0.0005,
\end{equation}
where $U = 0$ was fixed. To be conservative, we will require the contributions from the scalar sector alone to be smaller than the uncertainties. More precisely we will impose the NP contributions to satisfy $S_{NP} \leq 10^{-2}$ and $\Delta\rho_{NP} \leq 10^{-4}$.

Our results are shown in Fig.\,\ref{fig:scalar_sector_oblique_parameters}. We focus on the $\lambda_{21}$ versus $\kappa$ plane since $\lambda_{21}$ is directly related to the Charged-Higgs mass, while $\kappa$ controls the masses and mixing of all scalar states (see Sec.\,\ref{sec:model}). Given that the parameter space in the scalar sector is rather large, we fix the remaining parameters to the following benchmarks: 
\begin{itemize}
    \item $\lambda_{22} = \lambda_{SS} = 3 \pi$, $\lambda_{12} = \lambda_{1S} = \lambda_{2S} = 10^{-3}$, and $t_{\beta, \eta} = 5$;
    \item $\lambda_{22} = \lambda_{SS} = 1$, $\lambda_{12} = \lambda_{1S} = \lambda_{2S} = 10^{-3}$, $\tbeta = 5$, and $\teta = 15$.
\end{itemize}
With these choices, the singlet-like CP-even scalar has a mass around $1\,\text{TeV}$ and is thus decoupled\footnote{Such large quartic self-couplings were chosen to guarantee that the singlet-like Higgs is heavy compared to the electroweak scale, and the CP-even scalar coming from the extra doublet is heavier than the SM-like one. However, they also imply a Landau pole close to $\mathcal{O}(\text{TeV})$, yielding a theory that quickly becomes non-perturbative. In order for the Landau pole to appear at higher energies, we must either consider smaller values of $\lambda_{ii}$, or larger values of $\text{t}_{\eta}$. Since we want to avoid reducing the gauge sector in our model to the usual DP one, we would need to further decrease the quartic self-couplings and, consequently, have lighter CP-even scalars in the spectrum. In this scenario, it is not clear whether we can use the 2HDM approximation and, for that reason, one would have to fully analyze the loop contributions of the scalar sector to $S$, $T$, $U$.}. 
As in 2HDMs, electroweak observables do not provide any information on the absolute mass scale of the additional scalar bosons in our model \cite{GfitterStatistics}. However, they can constrain their relative masses -- this follows from the one-loop vacuum polarization diagrams contributing to the oblique parameters \cite{huntersguide, rhoParameter}. It has been shown that, as a consequence, a highly non-degenerate scalar spectrum can lead to large corrections to $S$, $T$, $U$ \cite{StUparameters}. Fig.\,\ref{fig:scalar_sector_oblique_parameters} therefore illustrates regions of the $\lambda_{21}$-$\kappa$ parameter space in which the charged-Higgs, the pseudoscalar, and one of the extra CP-even states are nearly degenerate in mass, even though the overall mass scale remains undetermined.

The regions excluded by the $S$ ($\Delta\rho$) parameter are shaded in light (dark) gray. As we can see from Fig.\,\ref{fig:scalar_sector_oblique_parameters}, only the region $50\,\text{GeV}\lesssim \kappa \lesssim 1\,\text{TeV}$ is still allowed -- unless $\lambda_{21} \gtrsim 0.2$, in which case the lower bound gradually increases to about $120\,\text{GeV}$. This region corresponds to pseudoscalar masses between a few and a few hundred GeV. Both $S$ and $\Delta \rho$ exclude regions with small and large $\kappa$. The region with large $\kappa$ is excluded because the mass of the SM-like scalar deviates substantially from 125 GeV -- see Eq.\,(\ref{eq:ScalarMatrices}). Since both $S$ and $\Delta\rho$ measure deviations from the SM, this effect turns out to be important. As for the lower bounds, they come from regions in which the spectrum of scalars heavier than the $Z$ boson is not degenerate. More precisely, $S$ excludes regions with large mass gaps between the second heaviest CP-even state and either pseudoscalar or charged Higgs, while $\Delta\rho$ excludes regions in which the large mass gap is between the pseudoscalar and charged states.

We thus conclude that, at least for the slice of parameter space shown, large regions in $\lambda_{21}$ and $\kappa$ are still allowed. Although a complete scan of the scalar parameter space is beyond the scope of this paper, we have analyzed a few benchmarks to understand whether this property persists for other choices of parameters. Since the ratio of the vevs controls important features of both scalar and gauge sectors, we have chosen benchmarks with fixed values of $\lambda_{ij}$ and $\kappa$, and varying $\tbeta$ and $\teta$. We find that, typically, the region $\tbeta \gtrsim 2$ is open for any value of $\teta$, almost independently from the choice of the other parameters. The bound essentially comes from the requirement that the coupling between the SM-like scalar and fermions is close to its SM value. As in the type I 2HDM, the Yukawa coupling scales as $1/\text{s}_{\beta}$, so that requiring it to be within about $10\%$ of the SM value implies $\tbeta \gtrsim 2$. A similar conclusion was reached in\,\cite{Davoudiasl:2012ag}. It is also interesting to understand what happens in the $\tbeta, \teta \to \infty$ limit in which we recover the usual DP model. In this limit, the lightest scalar aligns with the SM-like Higgs boson, while all the remaining scalars become heavy and almost degenerate \cite{HiggsDecoupling}. The contributions to EWPOs are negligible and no relevant bound emerges, in accordance with the results of\,\cite{curtin}.

Finally, we have analyzed whether collider limits can bound the parameter space in the scalar sector. We have considered searches for charged Higgs bosons produced in VBF processes and decaying into WZ pairs\,\cite{CMSsearch}, and searches for pseudoscalars produced via gluon-gluon fusion\,\cite{CMS:pseudoscalar}. To compare with data, we have computed the appropriate cross-sections and decay rates, finding that in our model they are orders of magnitude below the experimental bound, so that no limit emerges.

\subsection{Bounds on the Gauge Sector from EWPOs} \label{sec:fitGaugeSector}

\begin{figure}
    \centering
    \includegraphics[width=0.49\textwidth]{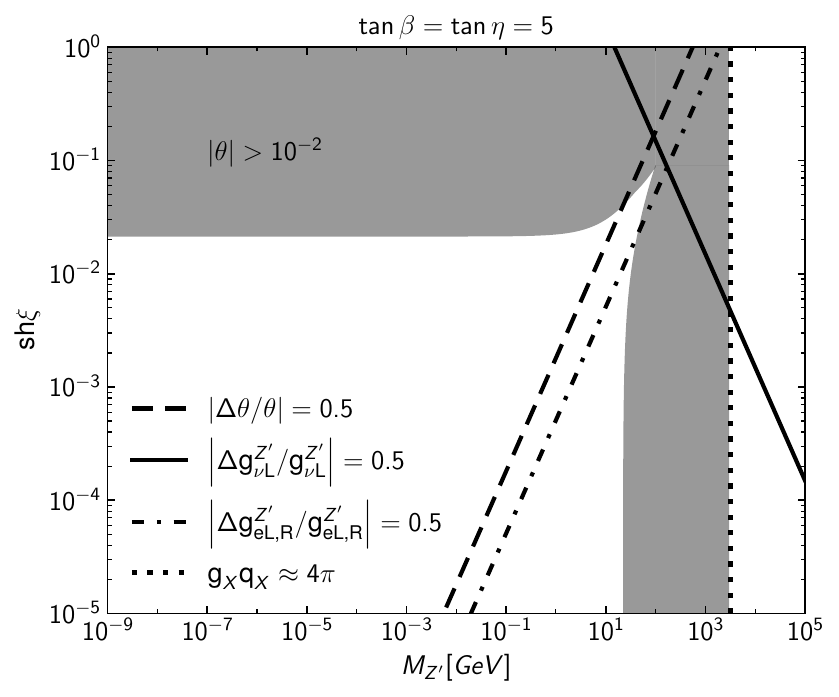}
    \includegraphics[width=0.49\textwidth]{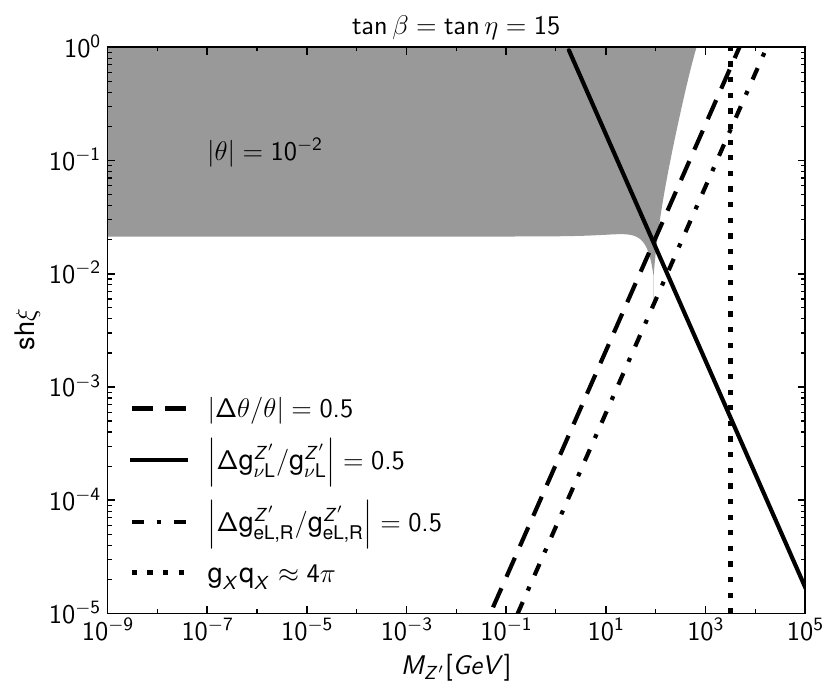}
    \caption{Regions in which major deformations with respect to the usual DP physics ($\text{t}_{\beta, \eta} \to \infty$) are expected, fixing $\tbeta = \teta = 5$ (left panel) and 15 (right panel). In the dark gray region, $\theta > 10^{-2}$ (i.e., we expect to have effects on EWPOs). Below the dashed line we have $\Delta\theta/\theta > 0.5$, while below the dot-dashed (continuous) line we have $\Delta g_{e_L}^{Z'}/g_{e_L}^{Z'}> 0.5$ ($\Delta g_{\nu_L}^{Z'}/g_{\nu_L}^{Z'}> 0.5$). The vertical dotted line shows the maximum allowed value for $M_{Z'}$, reached when $g_X q_X \simeq 4\pi$.}
    \label{fig:mod_angle_coupling}
\end{figure}

\begin{figure}[t]
    \centering
    \includegraphics[width=0.52\textwidth]{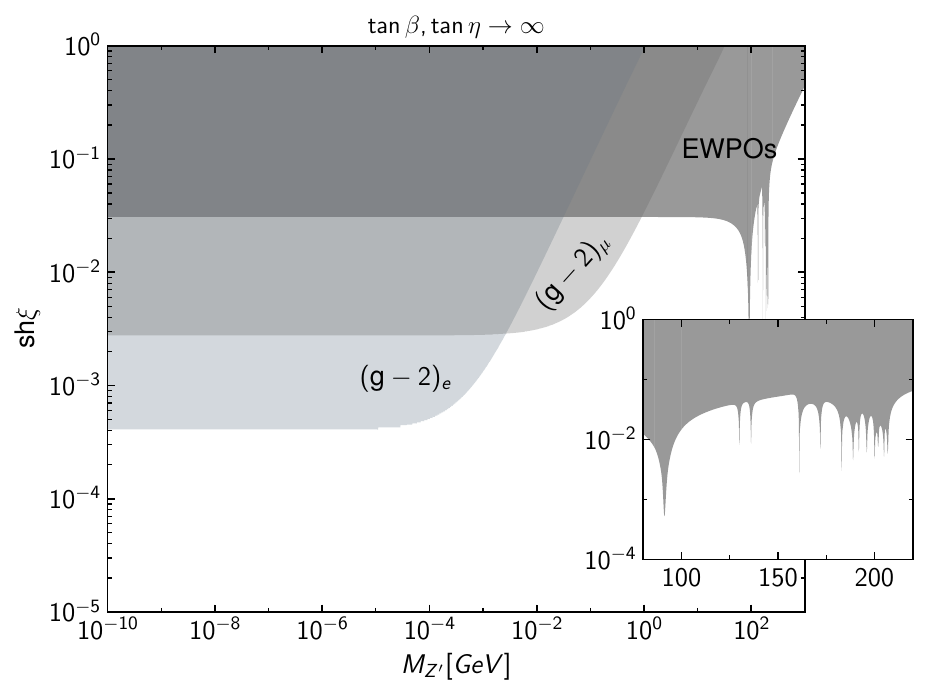} \\
    \includegraphics[width=0.49\textwidth]{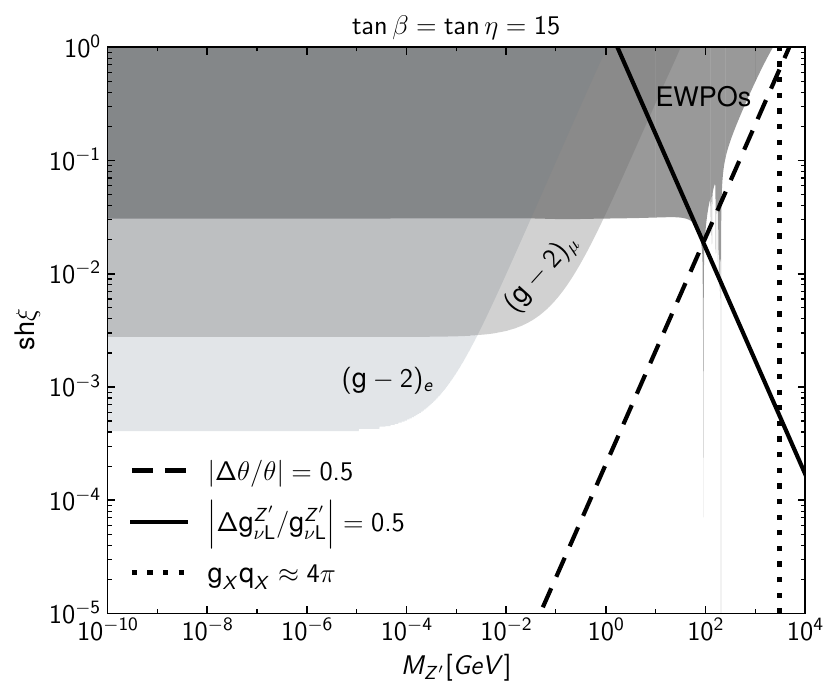}
    \includegraphics[width=0.49\textwidth]{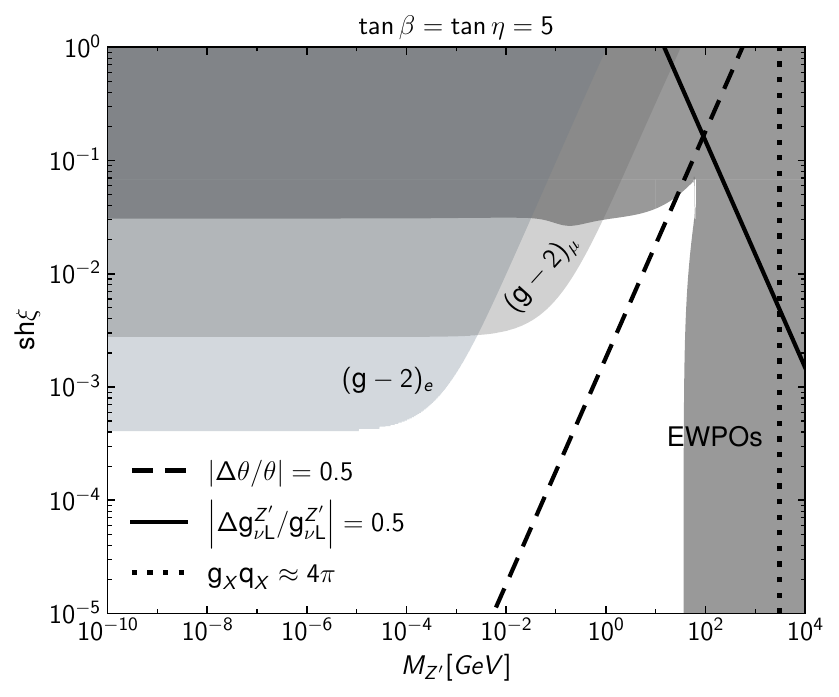}
    \caption{Current 68\% CL bounds on both kinetic mixing and DP mass. Constraints are derived from the fit to EWPOs (darkest gray shaded region) and lepton $g-2$ (electron bound is shown in light gray, while muon bound is shown in dark gray). The upper panel corresponds to the standard DP framework ($\text{t}_{\beta, \eta} \rightarrow \infty$ in our model). The lower panels present the excluded regions for a DP with generalized mixing: $\tbeta = \teta = 15$ (left) and $\tbeta = \teta = 5$ (right).}
    \label{fig:parameter_space}
\end{figure}

We now turn to the gauge sector of the model. Before discussing our main results, we ask in which regions of parameter space we may expect major differences between the limits of the generalized and usual DP (shown in the upper panel of Fig.\,\ref{fig:parameter_space}). We quantify the modifications using the relative variations of the mixing angle and $Z'$ couplings between the generalized DP (denoted by $\theta$, $g_\psi^{Z'}$) and the usual DP ($\theta_\text{DP}$, $g_{\psi,\text{DP}}^{Z'}$), defined as
\begin{equation}\label{eq:relative_variations}
    \frac{\Delta\theta}{\theta} \equiv \frac{\theta - \theta_\text{DP}}{\theta_\text{DP}}, ~~~~~~~~ \frac{\Delta g_{\psi}^{Z'}}{g_{\psi}^{Z'}} \equiv \frac{g_{\psi}^{Z'}- g_{\psi,\text{DP}}^{Z'}}{g_{\psi,\text{DP}}^{Z'}}.
\end{equation}
The results are shown in Fig.\,\ref{fig:mod_angle_coupling} for $\tbeta = \teta = 5$ (left panel) and 15 (right panel). In the dark gray region, we have $|\theta| > 10^{-2}$, such that modifications to EWPOs are expected. Below the dashed line we find $|\Delta\theta/\theta| > 0.5$, which corresponds to $\mathcal{O}(1)$ deformations in the $Z$ couplings of the generalized DP with respect to the usual DP. If the two regions intersect, we expect the EWPOs bounds to be different between the generalized and usual DP. We also show the lines below which the $Z'$ couplings to electrons (dot-dashed) and neutrinos (continuous) have $\mathcal{O}(1)$ differences with respect to the DP case. We do not show the same line for quarks since it is similar to the electron one. Finally, the vertical dotted line corresponds to the maximum allowed value for $M_{Z'}$, only reached when $g_X q_X \simeq 4\pi$.

As we see from Fig.\,\ref{fig:mod_angle_coupling}, for relatively small $\tbeta = \teta = 5$, the dark gray region extends below the dashed line, so that major modifications are expected in the limits coming from EWPOs. For larger values $\tbeta = \teta = 15$, this does not happen and we expect the limits to be essentially the same as in the usual DP model. Concerning the modifications to the $Z'$ couplings to electrons and quarks, they are expected for a relatively heavy $Z'$ and small $\sh$. The case of neutrinos is different, since major modifications are expected in most of the parameter space, especially for relatively light $Z'$. This is not surprising: in the usual case, a light DP couples to the electromagnetic current, i.e. there is no coupling between neutrinos and the $Z'$. This can be easily seen comparing Eq.\,\eqref{eq:ZprimeCouplings} with Eq.\,\eqref{eq:GaugeMixingAngle}: the neutrino coupling is $g_{\nu_L}^{Z'} = (\hat{s}_W \,\sh\,\cos\theta - \sin\theta)/2$, while in the $\text{t}_{\beta,\eta} \to \infty$ limit the mixing angle is approximately 
$\theta \simeq \hat{s}_W \,\sh$, resulting in $g_{\nu_L}^{Z'} \simeq 0$ to leading order in the angle. In our case, the correlation that leads to the cancellation breaks down due to the generalized mixing, so that even a light $Z'$ couples to neutrinos. 

We now turn to the actual excluded regions in parameter space. Our main results are summarized in Fig.\,\ref{fig:parameter_space}, in which we show the region excluded by EWPOs (shaded in dark gray), as well as the regions excluded by the low energy measurements of $(g-2)_\mu$ (medium gray) and $(g-2)_e$ (light gray). To draw the limits, we have used the procedure outlined in App.\,\ref{app:experimentalValues}. The different panels refer to different choices of $t_{\beta, \eta}$, taken to satisfy $\tbeta = \teta$ for simplicity:  $t_{\beta, \eta} \to \infty$ (usual DP case, upper panel), $\tbeta = \teta = 15$ (lower left), and $\tbeta = \teta = 5$ (lower right). Concerning the usual DP case (upper panel), we recover the known results\,\cite{curtin, hook,pospelov}. As we can see, the kinetic mixing is more strongly constrained by EWPOs when $M_{Z'} \approx M_Z$ or when $M_{Z'}$ approaches any of the LEP2 center of mass energies. Moreover, strong limits emerge from $(g-2)_{e,\mu}$ for a light $Z'$. As anticipated in Sec.\,\ref{sec:observables}, the difference in mass between electrons and muons ensures that the two limits are complementary.

We now move to the two lower panels in which the generalized mass mixing is turned on (i.e., $t_{\beta, \eta}$ are finite). In addition to the excluded regions (same color code as in the upper panel), and in order to make easier the comparison with Fig.\,\ref{fig:mod_angle_coupling}, we show where $\mathcal{O}(1)$ modifications to the mixing angle are expected (below the dashed line), as well as where modifications to the $Z'$ coupling to neutrinos will appear (below the continuous line). We further show a vertical dotted line corresponding to $g_X q_X \simeq 4\pi$ -- this is the point in which the naive perturbative limit of our theory and the $Z'$ mass cannot be pushed to larger values. Such vertical line is the same for both $\tbeta = \teta = 5$ and 15. This can be understood from Eqs.\,\eqref{SMzMass}-\eqref{eq:eigenmasses}: apart from relatively small corrections, we have 
\begin{equation}\label{eq:MZp_tbeta_teta}
M_{Z'} \simeq \hat{M}_X = 2 \frac{g_X\,q_X}{\sqrt{g^2+g'^2}} \sqrt{\frac{1+\teta^2}{1+\tbeta^2}} \,\hat{M}_Z, 
\end{equation}
so that for $\tbeta = \teta$, the dependence on the vevs ratios largely cancels and the maximum allowed value for $M_{Z'}$ is set solely by $g_{X} q_{X}$. 

Turning to the results, we observe that both panels follow exactly our discussion of Fig.\,\ref{fig:mod_angle_coupling}. For $\tbeta = \teta = 15$ (left), $\theta$ is so small in the region where the mass mixing dominates that no measurable effects on EWPOs are generated, and the exclusion is virtually undistinguishable from the usual DP case. On the other hand, for $\tbeta = \teta = 5$ (right), the generalized mass mixing excludes more parameter space with respect to the usual DP model. This happens precisely in the region where the mass mixing dominates and $\theta$ is sufficiently large to give an effect on EWPOs. In the right panel, we can also see a ``bump'' in the exclusion appearing for $M_{Z'} \simeq (0.1-0.2)\,\text{GeV}$. This is due to an enhancement of the $Z'$ contribution to $\nu-e$ scattering that happens for $M_{Z'} \sim \sqrt{m_{e} E_{\nu}}$. This bump is not present in the usual DP model because, as discussed above, a light DP does not couple to neutrinos. It is also not present in the lower left panel because, for $\tbeta = \teta =15$, the $Z'$ contribution to $\sigma(\nu e \to \nu e)$ is about one order of magnitude smaller than the experimental uncertainty, so that no signal appears. 

\begin{figure}
    \centering
    \includegraphics[width=0.49\linewidth]{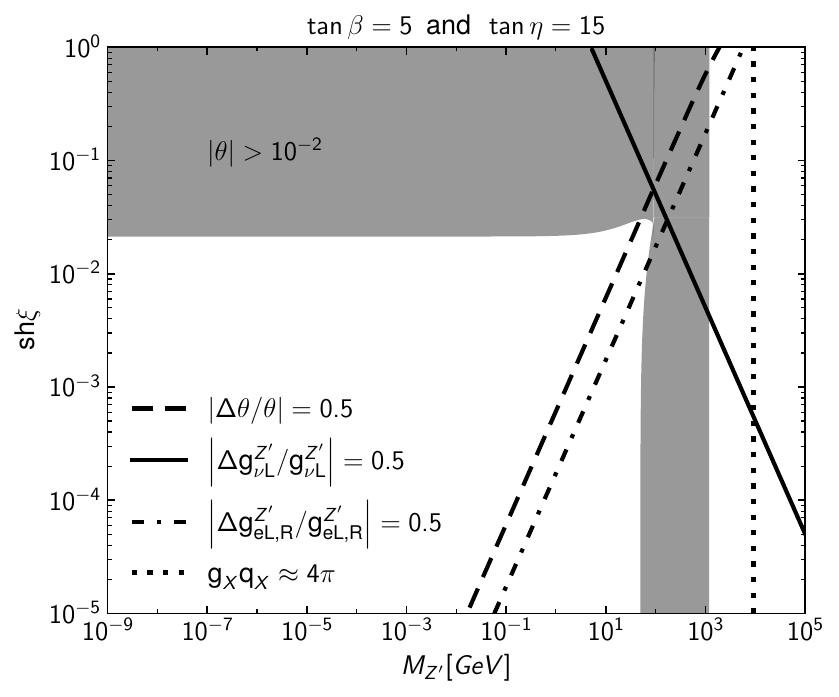}
    \includegraphics[width=0.49\linewidth]{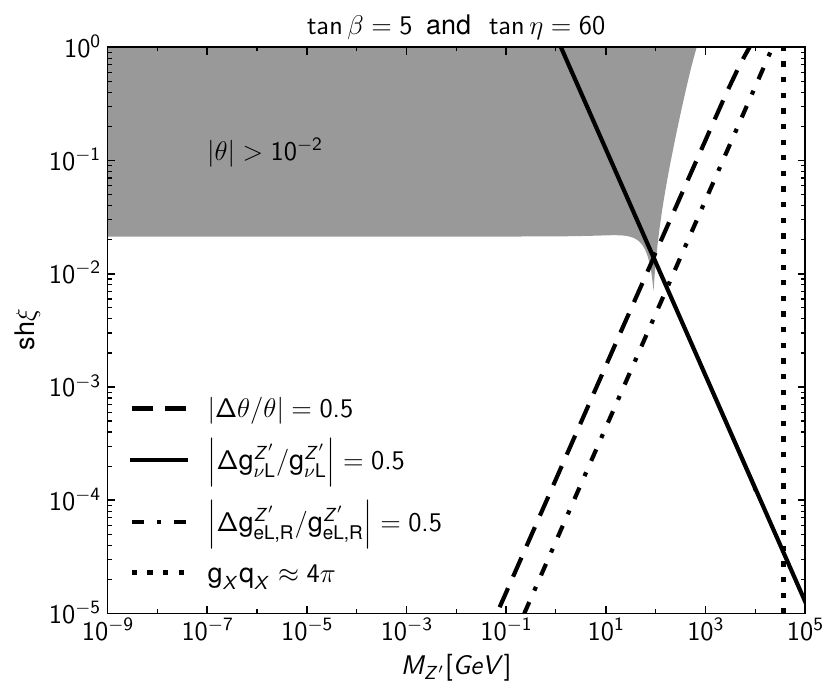}
    \caption{As in Fig.\,\ref{fig:mod_angle_coupling} but for $\tbeta =5$, $\teta = 15$ (left panel), and $\tbeta = 5$, $\teta = 60$ (right panel).}
    \label{fig:Z'_decoupling}
\end{figure}

For $\tbeta = \teta = 5$, the exclusion coming from EWPOs extends up to $M_{Z'} \simeq 1\,\text{TeV}$, the maximum allowed value for the $Z'$ mass. This is a consequence of the choice $\tbeta = \teta$. What happens if we relax this choice? As it is clear from Eq.\,\eqref{eq:MZp_tbeta_teta}, when $\teta \gg \tbeta$, we have $M_{Z'} \gg \hat{M}_Z \simeq M_Z$ and some form of decoupling is to be expected. We can make this statement more precise inspecting Eq.\,\eqref{eq:GaugeMixingAngle}. Observing that the combination $c_\eta\,\hat{M}_X$ does not depend on $\teta$, we see that in the $\teta \gtrsim \tbeta$ regime, the main contribution to the generalized mixing is given by the second term in the numerator of $\tan(2\theta)$, which only depends on $\tbeta$. As we increase $\teta$ and go into the $\teta \gg \tbeta$ regime, the term proportional to $(\hat{M}_X/\hat{M}_Z)^2$ in the denominator starts to dominate, so that we expect $|\tan2\theta| \propto \teta^{-2}$, corresponding to the expected decoupling. This point is quantified in Fig.\,\ref{fig:Z'_decoupling}, the analog of Fig.\,\ref{fig:mod_angle_coupling} in which we keep fixed $\tbeta = 5$ and increase $\teta$ to 15 (left panel) and 60 (right panel). In the left panel, $\teta$ is still sufficiently moderate that the mixing angle is dominated by the $\tbeta$ contribution and we still exclude masses in the $40\,\text{GeV} \lesssim M_{Z'} \lesssim 1\,\text{TeV}$ range. We start, however, to see some decoupling, since the region $1\,\text{TeV} \lesssim M_{Z'} \lesssim 10\,\text{TeV}$ in unconstrained. On the other hand, for $\teta = 60$ we are deep into the $\teta \gg \tbeta$ regime and the mixing angle is so small that there is no limit from EWPOs. In this case we expect to be, once more, back to the usual DP case.

\section{Conclusions} \label{sec:conclusions}
$U(1)_X$ models are among the simplest extensions of the SM of particle physics and have the nice property that they can be searched for in a plethora of different experiments, from colliders to beam dumps to precision low energy experiments. Among the most robust limits are those coming from EWPOs, expected to be significant since, in general, the gauge sector of the SM is modified in $U(1)_X$ models. DP models are particularly simple because, under the assumption that the only portal with the SM is given by the kinetic mixing and that the vector mass is of St\"uckelberg type, the phenomenology is completely fixed by only two parameters: the DP mass and the kinetic mixing itself. An interesting question that can be asked is: how robust are the DP limits to small deformations to this scenario? Steps in this direction have been taken in \cite{Barducci:2021egn,Foguel:2022unm}, in which the effects due to the addition of higher dimensional operators or of a dark Higgs boson have been studied. 

In this paper, we take a further step in the same direction: we study a situation where, in addition to the kinetic mixing parameter, a mass mixing between the DP and the $Z$ boson is also considered. First we identified a suitable model in which such a generalized mixing emerges (Sec.\,\ref{sec:model}). This turns out to require the addition of one singlet and one doublet to the scalar sector, both charged under $U(1)_X$. The gauge sector of the model is completely described by four parameters: the kinetic mixing, two ratios of vevs (that we call $\tbeta$ and $\teta$), and the $U(1)_X$ gauge coupling. One combination of the latter three parameters controls the $Z'$ mass, while another combination controls the additional source of mass mixing. 
Since any mass mixing between the $Z$ and $Z'$ modifies the electroweak sector of the SM, we have then focused on the limits arising from EWPOs, described in Sec.\,\ref{sec:EWPOs}. 

Our results are presented in Sec.\,\ref{sec:constraints}. Concerning the scalar sector, we identify regions in the parameter space allowed by current data for some given benchmarks. Concerning instead the gauge sector, the main conclusion is that, for moderate values of $\tbeta$ and $\teta$, EWPOs exclude a much larger region in parameter space with respect to the usual DP with only kinetic mixing. However, as soon as $\tbeta$ and/or $\teta$ become larger, the model becomes virtually indistinguishable from the usual DP model, at least for what concerns EWPOs. 

Of course, this does not mean that in this region of parameter space there are no differences between generalized and usual DP. As can be seen from Figs.\,\ref{fig:mod_angle_coupling} and \ref{fig:Z'_decoupling}, large modifications to the $Z'$ coupling to SM particles are expected even when there are no modification to EWPOs, given that the latter are mainly driven by the $Z$ coupling and mixing angle $\theta$. A simple computation shows that, for the total $Z'$ decay width $\Gamma_{Z'}$, we can have $\Gamma_{Z'} = \mathcal{O}(10) \,\Gamma_{Z', \text{DP}}$ (with $\Gamma_{Z',\text{DP}}$ the total $Z'$ decay width for the usual DP) right below the dashed lines in Figs.\,\ref{fig:mod_angle_coupling}-\ref{fig:Z'_decoupling}. In such region, we thus expect major modifications to the limits coming from experiments in which the $Z'$ is directly produced and further decays to generate a signal. Comparing with Fig.\,3.3 of \cite{dark_photons}, we see that modifications can be expected in the limits coming from hadronic and electronic beam dumps (nu-Cal, CHARM, E137), colliders (Babar, CMS, Atlas), and also in supernova bounds. We expect the modifications to the limits to be particularly important for beam dumps and supernova. In these cases, the $Z'$ should decay at a macroscopic distance from the production point, and our model predicts a lifetime which is much shorter than the usual one in a large part of the parameter space where such bounds apply. 

Consider, for example, beam-dump experiments in which a DP is searched for as a displaced vertex \cite{dark_photons}. It has been shown that, in the usual DP scenario, beam-dump experiments give the best sensitivity for a kinetic mixing in the $10^{-7}-10^{-3}$ range \cite{Blumlein:2013cua}. Moreover, in the $1\,\text{MeV}-1\,\text{GeV}$ mass range, the decay length is such that more decays should fall inside the experiment's fiducial region, providing a stringent exclusion in the usual DP parameter space \cite{dark_photons, Blumlein:2013cua}. In our model, however, a very long-lived DP (i.e., very small kinetic mixing), for which most decays would typically happen past the detector, now presents a much shorter decay length. As a consequence, the number of decays in the detector might increase, leading to even stronger constraints. On the other hand, as the kinetic mixing increases, the shortening of the decay length might instead move decays closer to the interaction point, reducing the decay probability in the fiducial volume. The effect could be a gain of parameter space when compared to the usual DP model.

Since a case-by-case analysis is needed to asses quantitatively how these bounds are modified, we defer such study to future work.

\acknowledgments

We thank Innes Bigaran for comments on the manuscript. FH is also grateful to Wan Zhen Chua, Ana Carolina Dantas, Heitor Ernandes, Geoffrey Fatin, Steven Ferrante, Denis Garcia, Margarita Gavrilova, Mitrajyoti Ghosh, Andrew Gomes, Ameen Ismail, Lillian Luo, Sridhar Prabhu, Yik Chuen San, and Namitha Suresh for the numerous conversations that direct or indirectly contributed to the completion of this work. The work of EB is partly supported by the Italian INFN program on Theoretical Astroparticle Physics (TAsP). CC and FH are supported in part by the NSF grant PHY-$2309456$. CC is supported in part by the US-Israeli BSF grant $2016153$.

\appendix
\section{Diagonalizing the Gauge and Scalar Sectors} \label{app:diagonalization}

\subsection{Gauge Sector} \label{app:gaugeSector}

We present here more details about the diagonalization of the gauge boson Lagrangian. Starting from Eq.\,\eqref{eq:kineticL}, the terms involving the gauge bosons are
\begin{equation} \label{eq:GaugeKineticLagrangian}
    \mathcal{L}_{\text{Kinetic}} \supset - \frac{1}{4} \hat{\boldsymbol{V}}^{T}_{\mu \nu} K \hat{\boldsymbol{V}}^{\mu \nu} + \frac{1}{2} \hat{\boldsymbol{V}}^{T}_{\mu} M^{2} \hat{\boldsymbol{V}}^{\mu},
\end{equation}
with fields before diagonalization defined as $\hat{\bm{V}} = (\hat{Z}, \hat{A}, \hat{X})^T$ and $\hat{V}_{\mu \nu}$ the corresponding field-strength tensor. The kinetic matrix receives contributions from the kinetic mixing parameter $\chi$ and is given by
\begin{equation} \label{eq:K-M2-matrices}
        K = \begin{pmatrix}
            1 & 0 & - \hats \, \chi \\
            0 & 1 & \hatc \, \chi \\
            - \hats \, \chi & \hatc \, \chi & 1
        \end{pmatrix}, 
\end{equation}
while the mass matrix is
\begin{equation}
        M^{2} = \begin{pmatrix}
            \frac{1}{4} (g^{2} + g'^{2}) (v_{1}^{2} + v_{2}^{2}) & 0 & - \frac{1}{2} g_{X} q_{X} \sqrt{(g^{2} + g'^{2})} v_{2}^{2} \\
            0 & 0 & 0 \\
            - \frac{1}{2} g_{X} q_{X} \sqrt{(g^{2} + g'^{2})} v_{2}^{2} & 0 & (g_{X} q_{X})^{2} (w^{2} + v_{2}^{2})
        \end{pmatrix}.
\end{equation}
To diagonalize this Lagrangian, we first diagonalize the kinetic terms applying the transformation \cite{burgess}
\begin{equation}
    \hat{\boldsymbol{V}} = L \tilde{\boldsymbol{V}} \equiv \begin{pmatrix}
        1 & 0 & \hats \, \sh \\
        0 & 1 & - \hatc \, \sh \\
        0 & 0 & \ch
    \end{pmatrix} \begin{pmatrix}
        \tilde{Z} \\
        \tilde{A} \\
        \tilde{X}
    \end{pmatrix}.
\end{equation}
The mass matrix in the tilded basis is simply given by $\tilde{M}^{2} = L^{T} M^{2} L$ and can be diagonalized by a rotation $R$ defined as
\begin{equation}
    \tilde{\boldsymbol{V}} = R \boldsymbol{V} \equiv \begin{pmatrix}
        \cos{\theta} & 0 & - \sin{\theta} \\
        0 & 1 & 0 \\
        \sin{\theta} & 0 & \cos{\theta}
    \end{pmatrix} \begin{pmatrix}
        Z \\
        A \\
        Z'
    \end{pmatrix},
\end{equation}
with the rotation angle $\theta$ shown in Eq.\,(\ref{eq:GaugeMixingAngle}) and physical masses given by
\begin{equation} \label{GaugeMasses}
        M_{Z, Z'}^{2} = \frac{1}{2} \left[ \text{Tr} \, \tilde{M}^{2}  \pm \text{sign} \left( 1 - \frac{\hat{M}_{X}^{2}}{\hat{M}_{Z}^{2}} \right) \sqrt{ ( \text{Tr} \, \tilde{M}^{2} )^{2} - 4 \det \tilde{M}^{2}} \right],
\end{equation}
where, for simplicity, we have defined
\begin{equation}
    \begin{aligned} \label{eq:trace_det}
        \text{Tr} \, \tilde{M}^{2} & = \hat{M}_{Z}^{2} \left[ 1 + \left( \ch \, \frac{\hat{M}_{X}}{\hat{M}_{Z}} \right)^{2} - 2 \, \hats \, \sh \, \ch \, \cbeta \, \ceta \frac{\hat{M}_{X}}{\hat{M}_{Z}} + \left( \hats \, \sh \right)^{2} \right], \\
        \det \tilde{M}^{2} & = \left[1 - (\cbeta \, \ceta)^{2} \right] \left( \ch \, \hat{M}_{X} \hat{M}_{Z} \right)^{2}.
    \end{aligned}
\end{equation}
The function sign(x) is used to ensure that $M_{Z} \rightarrow \hat{M}_{Z}$ and $M_{Z'} \rightarrow \hat{M}_{X}$ in the limit of no kinetic and mass mixing. 

\subsection{Scalar Sector} \label{app:scalarSector}

We now turn to the diagonalization of the scalar Lagrangian. We parametrize the scalar fields as
\begin{equation}
    H_1 = 
    \begin{pmatrix}
        \phi_1^+ \\
        \frac{v_1 + h_1 + i \phi_1^0}{\sqrt{2}}
    \end{pmatrix}, ~~~ 
    H_2 = 
    \begin{pmatrix}
        \frac{v_2 + h_2 + i \phi_2^0}{\sqrt{2}} \\
        \phi_2^-
    \end{pmatrix}, ~~~
    S = \frac{w + s + i \phi_S^0}{\sqrt{2}},
\end{equation}
where the vevs are determined minimizing the scalar potential of Eq.\,\eqref{eq:potential}:
\begin{equation}
    \begin{aligned}
        \mu_{1}^{2} & = \lambda_{11} v_{1}^{2} + \frac{\lambda_{12}}{2} v_{2}^{2} + \frac{\lambda_{1S}}{2} w^{2} + \frac{\kappa}{\sqrt{2}} \frac{v_{2} w}{v_{1}}, \\
        \mu_{2}^{2} & = \lambda_{22} v_{2}^{2} + \frac{\lambda_{12}}{2} v_{1}^{2} + \frac{\lambda_{2S}}{2} w^{2} + \frac{\kappa}{\sqrt{2}} \frac{v_{1} w}{v_{2}}, \\
        \mu_{S}^{2} & = \lambda_{SS} w^{2} + \frac{\lambda_{1S}}{2} v_{1}^{2} + \frac{\lambda_{2S}}{2} v_{2}^{2} + \frac{\kappa}{\sqrt{2}} \frac{v_{1} v_{2}}{w}.
    \end{aligned}
\end{equation}
These equations can be used to eliminate the mass parameters $\mu_i^2$, allowing to express the mass matrices in terms of the vevs. In the charged scalar sector we obtain
\begin{equation}
M_{\text{Charged}}^{2} = \frac{v^{2}}{2} \cbeta \begin{pmatrix} \lambda_{21} \, \cbeta + \sqrt{2} \, \frac{\kappa}{v} \frac{\teta}{\tbeta} & \lambda_{21} \, \sbeta + \sqrt{2} \, \frac{\kappa}{v} \teta \\
        \lambda_{21} \, \sbeta + \sqrt{2} \frac{\kappa}{v} \teta & \tbeta (\lambda_{21} \, \sbeta + \sqrt{2} \, \frac{\kappa}{v} \teta)    
        \end{pmatrix},
\end{equation}
in the CP-odd sector we have
\begin{equation}
    M_{\text{CP-odd}}^{2} = \frac{\kappa \, v}{\sqrt{2}} \cbeta \begin{pmatrix} \frac{\teta}{\tbeta} & \teta & 1 \\
        \teta & \tbeta \, \teta & \tbeta \\
        1 & \tbeta & \frac{\tbeta}{\teta}
        \end{pmatrix}, 
\end{equation}
and in the CP-even sector we have the matrix of Eq.\,\eqref{eq:ScalarMatrices}. Diagonalization can be performed with standard techniques (see, for instance, \cite{n2hdm} whose notation we follow). In the charged and CP-odd sectors, the rotation to the mass basis is simply given by
\begin{equation}
\begin{aligned}
    \begin{pmatrix}
        H^{\pm} \\
        G^{\pm}
    \end{pmatrix} & = 
    \begin{pmatrix} 
        \cbeta & \sbeta \\
        - \sbeta & \cbeta
    \end{pmatrix} 
    \begin{pmatrix}
        \phi_{1}^{\pm} \\
        \phi_{2}^{\pm}
    \end{pmatrix}, \\
    \begin{pmatrix}
        A^{0} \\
        G_{1}^{0} \\
        G_{2}^{0}
    \end{pmatrix} & = \frac{1}{\sqrt{\sbeta^{2} + \teta^{2}}} 
    \begin{pmatrix} 
        \teta & 0 & \sbeta \\
        0 & 1 & 0 \\
        - \sbeta & 0 & \teta
    \end{pmatrix} 
    \begin{pmatrix} 
        \cbeta & \sbeta & 0 \\
        - \sbeta & \cbeta & 0 \\
        0 & 0 & 1
    \end{pmatrix}
    \begin{pmatrix}
        \phi_{1}^{0} \\
        \phi_{2}^{0} \\
        \phi_{S}^{0}
    \end{pmatrix},
\end{aligned}
\end{equation}
where $H^\pm$ and $A^0$ are the physical states with masses shown in Eqs.\,\eqref{eq:ChargedMass}-\eqref{eq:OddMass}, and the states $G^\pm$ and $G_{1,2}^0$ are the would-be NGBs eaten up by the gauge bosons. Diagonalization of the CP-even mass matrix must instead be done numerically.  

\section{SM Predictions, Experimental Values of Observables, and Statistical \\Analysis} \label{app:experimentalValues}
\begin{table}[t]
    \centering
    \begin{tabular}{ c|c|c } 
        \hline
         Observable & Experiment & SM \\ 
        \hline
        $\Gamma_{Z}$ [GeV] & $2.4952 \pm 0.0023$ & $2.4942$ \\ 
        $\sigma_{\text{had}}$ [nb] & $41.540 \pm 0.037$ & $41.481$ \\
        $R_{l}$ & $20.767 \pm 0.025$ & $20.739$ \\    
        $\mathcal{A}_{FB}^{l}$ & $0.0171 \pm 0.0010$ & $0.01642$ \\
        $\mathcal{A}_{\tau}(P_{\tau})$ & $0.1439 \pm 0.0043$ & $0.1469$ \\ 
        $\mathcal{A}_{e}(P_{\tau})$ & $0.1498 \pm 0.0049$ & $0.1469$ \\ 
        $\mathcal{A}_{LR}^{e}$ & $0.15138 \pm 0.00216$ & $0.1469$ \\ 
        $R_{b}$ & $0.21629 \pm 0.00066$ & $0.21582$ \\ 
        $R_{c}$ & $0.1721 \pm 0.0030$ & $0.17221$ \\
        $\mathcal{A}_{FB}^{b}$ & $0.0992 \pm 0.0016$ & $0.01030$ \\ 
        $\mathcal{A}_{FB}^{c}$ & $0.0707 \pm 0.0035$ & $0.0735$ \\
        \hline
    \end{tabular}
    \caption{\label{tab:measurements1} Main electroweak observables measured at the $Z$ pole \cite{lepcollab} and their SM predictions \cite{pdg}.}
\end{table}

\begin{table}[t]
    \centering
    \begin{tabular}{ c|c|c|c|| c| c|c } 
        \hline
       Obs. & $\sqrt{s}$ [GeV] & Exp. [pb] & SM [pb] & $\sqrt{s}$ [GeV] & Exp. [pb] & SM [pb] \\ 
        \hline
       $\sigma (q \bar{q})$ &  130 &  $82.1 \pm 2.2$ & $82.8$ & 192 & $22.05 \pm 0.53$ & $21.24$ \\
       $\sigma (\mu^{+} \mu^{-})$ & 130 &  $8.62 \pm 0.68$ & $8.44$ & 192 &  $2.92 \pm 0.18$ & $3.10$ \\
       $\sigma (\tau^{+} \tau^{-})$ & 130 &  $9.02 \pm 0.93$ & $8.44$ & 192 & $2.81 \pm 0.23$ & $3.10$ \\
        \hline
       $\sigma (q \bar{q})$ & 136 &  $66.7 \pm 2.0$ & $66.6$ & 196 &  $20.53 \pm 0.34$ & $20.13$ \\
       $\sigma (\mu^{+} \mu^{-})$ &  136 & $8.27 \pm 0.67$ & $7.28$ & 196 &  $2.94 \pm 0.11$ & $2.96$ \\
       $\sigma (\tau^{+} \tau^{-})$ & 136 &  $7.078 \pm 0.820$ & $7.279$ & 196 &  $2.94 \pm 0.14$ & $2.96$ \\ 
        \hline
       $\sigma (q \bar{q})$ & 161 &  $37.0 \pm 1.1$ & $35.2$ & 200 &  $19.25 \pm 0.32$ & $19.09$ \\
       $\sigma (\mu^{+} \mu^{-})$ & 161 &  $4.61 \pm 0.36$ & $4.61$ & 200 &  $3.02 \pm 0.11$ & $2.83$ \\
       $\sigma (\tau^{+} \tau^{-})$ & 161 &  $5.67 \pm 0.54$ & $4.61$ & 200 &  $2.90 \pm 0.14$ & $2.83$ \\
        \hline
       $\sigma (q \bar{q})$ & 172 &  $29.23 \pm 0.99$ & $28.74$ & 202 &  $19.07 \pm 0.44$ & $18.57$ \\
       $\sigma (\mu^{+} \mu^{-})$ & 172 &  $3.57 \pm 0.32$ & $3.95$ & 202 &  $2.58 \pm 0.14$ & $2.77$ \\
       $\sigma (\tau^{+} \tau^{-})$ & 172 &  $4.01 \pm 0.45$ & $3.95$ & 202 &  $2.79 \pm 0.20$ & $2.77$ \\
        \hline
       $\sigma (q \bar{q})$ & 183 &  $24.59 \pm 0.42$ & $24.20$ & 205 &  $18.17 \pm 0.31$ & $17.81$ \\
       $\sigma (\mu^{+} \mu^{-})$ & 183 &  $3.49 \pm 0.15$ & $3.45$ & 205 &  $2.45 \pm 0.10$ & $2.67$ \\
       $\sigma (\tau^{+} \tau^{-})$ & 183 &  $3.37 \pm 0.17$ & $3.45$ & 205 & $2.78 \pm 0.14$ & $2.67$ \\
        \hline
       $\sigma (q \bar{q})$ & 189 &  $22.47 \pm 0.24$ & $22.156$ & 207 &  $17.49 \pm 0.26$ & $17.42$ \\
       $\sigma (\mu^{+} \mu^{-})$ & 189 &  $3.123 \pm 0.076$ & $3.207$ & 207 &  $2.595 \pm 0.088$ & $2.623$ \\
       $\sigma (\tau^{+} \tau^{-})$ & 189 &  $3.20 \pm 0.10$ & $3.20$ & 207 & $2.53 \pm 0.11$ & $2.62$ \\
    \end{tabular}
    \caption{\label{tab:measurements2} Cross sections for $e^{+} e^{-} \rightarrow \psi \bar{\psi}$ (with $\psi = q, \mu, \tau$) measured above the $Z$ resonance and their SM predictions \cite{lepcollab2}.}
\end{table}
\begin{table}
    \begin{tabular}[t]{c|c c c c}
        \hline
        & $\Gamma_{Z}$ & $\sigma_{\text{had}}$ & $R_{l}$ & $A_{FB}^{l}$ \\
        \hline
        $\Gamma_{Z}$ & $1.000$ & & & \\
        $\sigma_{\text{had}}$ & $-0.297$ & $1.000$ & & \\
        $R_{l}$ & $0.004$ & $0.183$ & $1.000$ & \\
        $A_{FB}^{l}$ & $0.003$ & $0.006$ & $-0.056$ & $1.000$ \\
        \hline
    \end{tabular} %
    \hspace{\fill}
    \begin{tabular}[t]{c|c c c c}
        \hline 
        & $R_{b}$ & $R_{c}$ & $A_{FB}^{b}$ & $A_{FB}^{c}$ \\
        \hline
        $R_{b}$ & $1.00$ & & & \\
        $R_{c}$ & $-0.18$ & $1.00$ & & \\
        $A_{FB}^{b}$ & $-0.10$ & $0.04$ & $1.00$ & \\
        $A_{FB}^{c}$ & $0.07$ & $-0.06$ & $0.15$ & $1.00$ \\
        \hline
    \end{tabular}
    \caption{\label{tab:correlations1} Correlation matrices among observables measured at the $Z$ peak \cite{lepcollab}.}
\end{table}
\begin{table}
    \centering
    \begingroup
    \setlength{\tabcolsep}{4pt}
    \begin{tabular}[t]{c|c c c c c c c c c c c c}
        \hline
        $\sqrt{s}$ & 130 & 136 & 161 & 172 & 183 & 189 & 192 & 196 & 200 & 202 & 205 & 207 \\
        \hline
        130 & 1.000 & & & & & & & & & & & \\
        136 & 0.071 & 1.000 & & & & & & & & & & \\
        161 & 0.080 & 0.075 & 1.000 & & & & & & & & & \\
        172 & 0.072 & 0.067 & 0.077 & 1.000 & & & & & & & & \\
        183 & 0.114 & 0.106 & 0.120 & 0.108 & 1.000 & & & & & & & \\
        189 & 0.146 & 0.135 & 0.153 & 0.137 & 0.223 & 1.000 & & & & & & \\
        192 & 0.077 & 0.071 & 0.080 & 0.072 & 0.117 & 0.151 & 1.000 & & & & & \\
        196 & 0.105 & 0.097 & 0.110 & 0.099 & 0.158 & 0.206 & 0.109 & 1.000 & & & & \\
        200 & 0.120 & 0.110 & 0.125 & 0.112 & 0.182 & 0.235 & 0.126 & 0.169 & 1.000 & & & \\
        202 & 0.086 & 0.079 & 0.090 & 0.081 & 0.129 & 0.168 & 0.090 & 0.122 & 0.140 & 1.000 & & \\
        205 & 0.117 & 0.109 & 0.124 & 0.111 & 0.176 & 0.226 & 0.118 & 0.162 & 0.184 & 0.132 & 1.000 & \\
        207 & 0.138 & 0.128 & 0.145 & 0.130 & 0.208 & 0.268 & 0.138 & 0.190 & 0.215 & 0.153 & 0.213 & 1.000 \\
        \hline
    \end{tabular}
    \endgroup
    \caption{\label{tab:correlations2} Correlation matrix among the hadronic cross sections $\sigma(q\bar{q})$ measured above the $Z$ peak for different energies \cite{lepcollab2}.}
\end{table}
\begin{table}[t]
    \centering
    \begin{tabular}{ c|c|c } 
        \hline
        Observable & Experiment & Standard Model \\ 
        \hline
        $g_{e V}$ & $-0.025 \pm 0.020$ & $-0.0398$ \\ 
        $g_{e A}$ & $-0.503 \pm 0.017$ & $-0.5064$ \\
        \hline
    \end{tabular}
    \caption{\label{tab:measurementsCHARM} Effective vector and axial coupling constants from neutrino-electron scattering measured by the CHARM II collaboration \cite{charm2}. The SM prediction for such couplings is given by PDG \cite{pdg}.}
\end{table}
For completeness, in this appendix we present the experimental data and SM predictions we have used to produce Fig.\,\ref{fig:parameter_space}. Table\,\ref{tab:measurements1} summarizes the experimental measurements and SM predictions for observables measured at the $Z$ peak, while in Table\,\ref{tab:measurements2} we show the same information for the measurement of the cross-section for the process $e^+e^- \to \psi \bar{\psi}$ (with $\psi = q, \mu, \tau$) performed above the $Z$ peak. Since for our statistical analysis we need the correlations between observables, we present them in Tables\,\ref{tab:correlations1} and \ref{tab:correlations2} for the observables at and above the $Z$ peak, respectively. 
In Table\,\ref{tab:measurementsCHARM} we show the values of the vector and axial couplings derived from neutrino-electron scattering measured at the CHARM II experiment \cite{charm2}, together with their SM prediction extracted from\,\cite{pdg}. Finally, in our analysis we also use the $W$ mass obtained by the ATLAS collaboration, $M_{W} = (80.370 \pm 0.019)$ GeV \cite{ATLASwMass}, with SM prediction of $80.358$ GeV \cite{pdg}.

To draw the exclusion region showed in Fig.\,\ref{fig:parameter_space}, we proceed as follows. The gauge sector of our model depends on four parameters, ($M_{Z'}$, $\sh$, $\tbeta$, $\teta$). Assuming the measurements to be normally distributed, we can define the chi-square as \cite{curtin}
\begin{equation}
    \begin{split}
        \chi^{2} (M_{Z'}&, \sh, \tbeta, \teta) = \\ 
        &\left[ x_{exp} - x_{theo} (M_{Z'}, \sh, \tbeta, \teta) \right]_{i} \left( cov^{-1} \right)_{ij} \left[ x_{exp} - x_{theo} (M_{Z'}, \sh, \tbeta, \teta) \right]_{j},
    \end{split}
\end{equation}
with covariance matrix given by $\left( cov \right)_{ij} = \sigma_{i} \left( cor \right)_{ij} \sigma_{j}$, where $\sigma$ is the standard deviation of each measurement and $cor$ the correlation matrix presented in Tables\,\ref{tab:correlations1} and \ref{tab:correlations2}. The exclusion region is obtained requiring\,\cite{GfitterStatistics} that
\begin{equation}
\Delta \chi^{2}(M_{Z'}, \sh, \tbeta, \teta) = \chi^{2} (M_{Z'}, \sh, \tbeta, \teta) - \chi^{2}_{min} \gtrsim 4.7,
\end{equation} 
which corresponds to an exclusion at the  $68 \%$ C.L. for four degrees of freedom. The quantity $\chi^{2}_{min}$ denotes the absolute minimum value of the $\chi^{2}$ function, obtained by comparing the SM predictions to the corresponding measurements.

\bibliographystyle{JHEP2}
\bibliography{biblio.bib}

\end{document}